\documentclass[11pt, letterpaper]{article}
\pdfoutput=1

\usepackage[LGR,T1]{fontenc}
\usepackage[utf8]{inputenc}
\usepackage{textcomp} 
\usepackage{amsmath,amssymb}  
\usepackage{eucal}  
\usepackage{amsthm}
\usepackage{bm}      
\usepackage{dsfont}  

\usepackage[style=authoryear]{biblatex}
\usepackage{geometry}
\usepackage{titlesec}
\titleformat{\section}
  {\normalfont\Large\bfseries}{\thesection}{10pt}{}
\titlespacing*{\section}
  {0pt}{0pt}{5pt}
\titleformat{\subsection}
  {\normalfont\large\bfseries}{\thesubsection}{8pt}{}
\titlespacing*{\subsection}
  {0pt}{0pt}{5pt}

\usepackage{verbatim,upquote}
\usepackage[justification=centering]{caption}
\usepackage{graphicx}
\usepackage[rgb,table, fixpdftex, hyperref]{xcolor}
\usepackage{listings}
\usepackage{afterpage}   
\usepackage{paralist}    
\usepackage{varwidth,pbox}  
\usepackage{float}
\usepackage{array}
\usepackage{calc}
\usepackage{dcolumn}
\usepackage{hhline}
\usepackage{textcomp} 
\usepackage{exscale,relsize}
  \newcommand{\lgsig}{{\mathlarger{\mathlarger{\mathlarger{\sigma}}}}}
\usepackage{blkarray}
\usepackage{mathtools}  

\usepackage{tikz}
  \usetikzlibrary{arrows}
  \usetikzlibrary{matrix}
  \usetikzlibrary{calc}

\theoremstyle{definition}
\newtheorem{defn}{Definition}
\theoremstyle{remark}
\newtheorem*{rmk*}{Remark}
\newtheorem*{note*}{Note}

\floatstyle{plaintop}
\newfloat{matlab}{ht}{matlab}
\floatname{matlab}{MATLAB Example}
\newfloat{algorithm}{ht}{algorithm}
\floatname{algorithm}{Algorithm}

\newcommand{\ticker}[1]{\text{\small\textsf{#1}}}
\newcommand{\smtick}[1]{\text{\footnotesize\textsf{#1}}}  
\newcommand{\sstick}[1]{\scalebox{.7}{\textsf{#1}}} 

\newlength{\pointradius}
\newlength{\smallradius}

\numberwithin{equation}{section}
\numberwithin{table}{section}
\numberwithin{figure}{section}
\numberwithin{algorithm}{section}

\colorlet{algcomment}{magenta!60!blue}
\definecolor{grid}{rgb}{0.7,0.7,0.7}
\colorlet {FBTv}{cyan!60!blue}
\colorlet {XBIv}{green}
\colorlet {UIPv}{yellow}
\colorlet {ZNSv}{violet!50}
\colorlet {CRPv}{red!70}
\colorlet {FBTa}{blue!70!black}
\colorlet {XBIa}{green!40!black}
\colorlet {UIPa}{brown!75!black}
\colorlet {ZNSa}{violet!95!black}
\colorlet {CRPa}{red!75!black}
\colorlet {XF}{blue!40!black}  
\colorlet {FZ}{violet!50!black}  
\colorlet {CR}{red!60!black}  
\colorlet {CRi}{red!15}

\usepackage[colorlinks=true,%
   citecolor=green!50!black,
   linkcolor=red!80!black,  
   urlcolor=blue!90!black,  
]{hyperref}

\begin{document}

\title{How Not To Do Mean-Variance Analysis}
\author{Vic Norton\\
  Mathematical Ruminations Inc\\
  622 Morton Avenue\\
  Bowling Green, OH 43402-2223\\
  \url{mailto:vic@norton.name}}
\date{October 22, 2018}
\maketitle
\thispagestyle{empty}

\begin{center}
\begin{minipage}{8cm}
\begin{abstract}\noindent
We use the 2014 market history of two high-returning biotechnology
exchange-traded funds to illustrate how ex post mean-variance analysis
should not be done. Unfortunately, the way it should not be done is the
way it generally is done---to our knowledge.
\end{abstract}
\end{minipage}
\end{center}

\newpage
\begin{center}
\includegraphics[scale=.25]{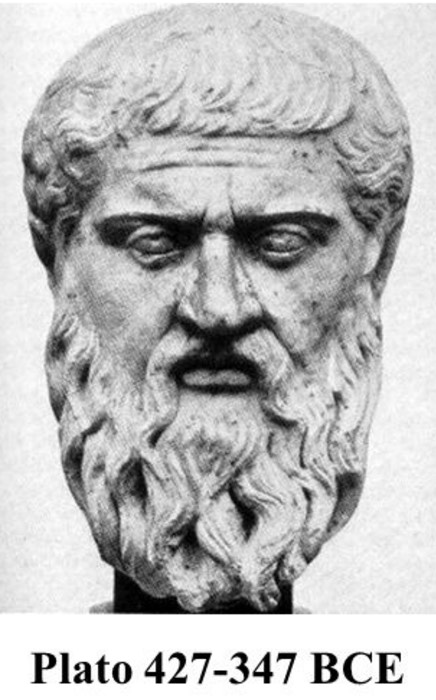}\\
\large\bfseries
\fontencoding{LGR}\selectfont%
>agewm`etrhtos\\
mhde`is\\
e>is'itw\\[1.0ex]
\large
\mdseries
\fontencoding{T1}\selectfont
Let no one ignorant of geometry enter here
\end{center}

\section{Preface}\label{preface}
Ex post mean-variance analysis is a financial application of
descriptive statistics. But descriptive statistics, where the sum of
square deviations from the mean plays a fundamental role, has a strong
geometric flavor. In this paper we emphasize the geometry of
mean-variance analysis.

Geometry starts with points.
The primary points in our geometric exposition are two biotechnology
exchange-traded funds,\\[1ex]\hspace*{1cm}%
  \ticker{FBT} -- First Trust NYSE Arca Biotechnology Indx Fund\\
and\\\hspace*{1cm}%
  \ticker{XBI}\hspace{0.7ex} -- SPDR S\&P Biotech ETF.\\[1ex]
The graphs of the 2013-12-31-normalized adjusted closing prices,
$\mathbf{a}_{\sstick{FBT}}$ and $\mathbf{a}_{\sstick{XBI}}$,
 of the two funds are shown in Figure \ref{FXUCplot}, as well as the
 graph of an unattended long portfolio, \ticker{UIP}, in the two funds.
 \ticker{UIP}'s normalized adjusted closing price vector,\\[-2ex]
\begin{equation}\label{convex_comb}
  \mathbf{a}_{\sstick{UIP}} = 0.75 \cdot \mathbf{a}_{\sstick{FBT}}
  + 0.25 \cdot \mathbf{a}_{\sstick{XBI}},
\end{equation}
is necessarily a convex combination of
$\mathbf{a}_{\sstick{FBT}}$ and $\mathbf{a}_{\sstick{XBI}}$.

Normalized adjusted closing prices are the horses that drive our
mean-variance cart. When the adjusted closing price vectors
~$\mathbf{a}_{\sstick{FBT}}, \mathbf{a}_{\sstick{XBI}}$, and
$\mathbf{a}_{\sstick{UIP}}$~
are replaced by their daily return vectors
~$\mathbf{r}_{\sstick{FBT}}, \mathbf{r}_{\sstick{XBI}}$, and
$\mathbf{r}_{\sstick{UIP}}$,~
you move into a higher dimensional space with all geometry left behind.

The pictures that follow tell the whole story. Thanks
to the \href{https://en.wikipedia.org/wiki/PGF/TikZ}{TikZ}
vector-graphics language these pictures are precise, numerical
images---not just schematic drawings.

\newpage
\section{Normalized adjusted closing prices -- a geometric example}
\label{FXUCexample}

\begin{figure}[H]
  \captionsetup{width=10cm}
  \caption{\label{FXUCplot}%
    2013-12-31-normalized adjusted closing prices of two biotechnology
    ETFs and two ``portfolios'' in these ETFs
  }
  \centering
  \vspace*{-2ex}
\begin{tikzpicture}[scale=1.1, >={angle 60},
  xscale=0.047,yscale=0.14]   
  
  \foreach \y in {100,110,120,130,140,150}
    \draw[grid] (0,\y) -- (252,\y);
  \foreach \y in {90,100,110,120,130,140,150,160}
    \draw (252,\y) -- (255,\y) node[right]{\y};
  \foreach \x in {61,124,188}
    \draw[grid] (\x,90) -- (\x,160);
  \foreach \x in {0,61,124,188,252}
    \draw (\x,90) -- (\x,89);
  \node[below=1] at (0,89) {2013-12-31};
  \node[below=1] at (61,89) {2014-03-31};
  \node[below=1] at (124,89) {2014-06-30};
  \node[below=1] at (188,89) {2014-09-30};
  \node[below=1] at (252,89) {2014-12-31};
  \draw[thick] (0,90) rectangle (252,160);
  
  \filldraw[fill=white] (4.7,142.5) rectangle (200,157);  
  
  \draw[line width=3pt,FBTa] (15,154.5) -- (20,154.5)
    node[black,right=2]{\ticker{FBT} =
    First Trust Biotechnology Index ETF};
  \draw[line width=3pt,XBIa] (15,151.5) -- (20,151.5)
    node[black,right=2]{\ticker{XBI} \hspace*{0.7ex}=
    SPDR S\&P Biotech ETF};
  \draw[line width=3pt,UIPa] (15,148.5) -- (20,148.5)
    node[black,right=2]{\ticker{UIP} \hspace*{0.7ex}=
    Unattended \ticker{FBT}-\ticker{XBI} investment portfolio};
  \draw[line width=3pt,CRPa] (15,145.2) -- (20,145.2)
    node[black,right=2]{\ticker{CRP} \hspace*{0.1ex}=
    Continually reallocated \ticker{FBT}-\ticker{XBI} portfolio};
  
  \draw[CRPa,line width=1pt] plot coordinates {  (0,100.00) 
  (1,100.45) (2,99.94) (3,99.41) (4,101.34) (5,103.11)
  (6,106.04) (7,109.12) (8,106.88) (9,110.49) (10,110.80)
  (11,112.46) (12,112.65) (13,115.16) (14,114.67) (15,114.86)
  (16,110.77) (17,107.81) (18,110.13) (19,109.34) (20,112.36)
  (21,110.47) (22,106.41) (23,107.35) (24,106.07) (25,105.61)
  (26,110.20) (27,112.64) (28,114.08) (29,114.50) (30,115.81)
  (31,114.59) (32,117.60) (33,115.78) (34,118.50) (35,120.36)
  (36,121.69) (37,127.64) (38,125.65) (39,125.38) (40,121.99)
  (41,122.44) (42,125.06) (43,124.99) (44,121.69) (45,121.68)
  (46,122.19) (47,120.85) (48,121.70) (49,119.03) (50,118.52)
  (51,118.67) (52,122.74) (53,121.49) (54,120.83) (55,115.94)
  (56,112.49) (57,112.35) (58,109.74) (59,111.51) (60,107.13)
  (61,110.67) (62,113.36) (63,113.43) (64,109.86) (65,106.15)
  (66,106.33) (67,105.90) (68,109.85) (69,103.40) (70,100.34)
  (71,99.72) (72,100.41) (73,103.11) (74,102.81) (75,104.38)
  (76,107.89) (77,105.87) (78,105.35) (79,101.64) (80,101.65)
  (81,104.88) (82,105.67) (83,106.54) (84,104.87) (85,107.39)
  (86,105.58) (87,104.40) (88,102.04) (89,104.41) (90,107.33)
  (91,106.24) (92,106.70) (93,104.90) (94,104.44) (95,106.64)
  (96,104.61) (97,105.02) (98,106.89) (99,106.75) (100,109.34)
  (101,108.93) (102,109.48) (103,108.72) (104,108.11) (105,108.85)
  (106,110.55) (107,112.01) (108,112.17) (109,114.59) (110,115.55)
  (111,115.25) (112,115.47) (113,115.77) (114,116.73) (115,116.38)
  (116,117.37) (117,117.38) (118,118.21) (119,117.28) (120,118.07)
  (121,118.20) (122,118.19) (123,118.74) (124,119.20) (125,121.72)
  (126,121.94) (127,122.26) (128,119.30) (129,116.42) (130,118.03)
  (131,117.20) (132,118.45) (133,118.90) (134,115.98) (135,114.59)
  (136,111.40) (137,114.52) (138,114.61) (139,115.55) (140,119.45)
  (141,117.75) (142,116.89) (143,115.87) (144,119.00) (145,119.61)
  (146,116.40) (147,116.36) (148,117.51) (149,117.82) (150,118.18)
  (151,117.04) (152,118.63) (153,119.91) (154,119.49) (155,121.99)
  (156,123.48) (157,123.87) (158,125.08) (159,125.42) (160,125.47)
  (161,123.97) (162,124.74) (163,129.66) (164,131.49) (165,131.13)
  (166,130.22) (167,132.19) (168,132.04) (169,132.16) (170,130.23)
  (171,129.97) (172,131.34) (173,129.59) (174,132.53) (175,131.72)
  (176,130.31) (177,127.99) (178,129.77) (179,130.50) (180,131.22)
  (181,130.73) (182,129.54) (183,128.94) (184,132.72) (185,130.20)
  (186,131.50) (187,131.92) (188,129.93) (189,128.16) (190,128.31)
  (191,130.39) (192,128.31) (193,125.69) (194,129.08) (195,125.84)
  (196,125.33) (197,123.14) (198,122.92) (199,124.45) (200,126.78)
  (201,127.69) (202,129.36) (203,132.08) (204,131.59) (205,135.35)
  (206,137.02) (207,138.13) (208,140.12) (209,139.08) (210,141.53)
  (211,141.34) (212,141.63) (213,140.25) (214,137.54) (215,139.87)
  (216,138.21) (217,140.37) (218,141.02) (219,141.38) (220,140.64)
  (221,137.56) (222,137.88) (223,140.40) (224,139.84) (225,140.63)
  (226,141.62) (227,144.26) (228,144.10) (229,145.82) (230,145.68)
  (231,143.49) (232,146.05) (233,146.17) (234,145.22) (235,146.62)
  (236,149.11) (237,150.63) (238,147.51) (239,147.66) (240,146.18)
  (241,141.47) (242,139.89) (243,145.31) (244,150.13) (245,151.58)
  (246,149.99) (247,143.11) (248,145.38) (249,148.90) (250,149.38)
  (251,147.59) (252,147.32)
  };
  
  \draw[UIPa,line width=1pt] plot coordinates {  (0,100.00) 
  (1,100.45) (2,99.94) (3,99.41) (4,101.34) (5,103.11)
  (6,106.05) (7,109.18) (8,106.94) (9,110.48) (10,110.80)
  (11,112.47) (12,112.67) (13,115.18) (14,114.69) (15,114.88)
  (16,110.79) (17,107.79) (18,110.13) (19,109.32) (20,112.34)
  (21,110.43) (22,106.36) (23,107.30) (24,106.01) (25,105.55)
  (26,110.15) (27,112.61) (28,114.04) (29,114.46) (30,115.77)
  (31,114.54) (32,117.55) (33,115.72) (34,118.45) (35,120.32)
  (36,121.65) (37,127.54) (38,125.58) (39,125.31) (40,121.90)
  (41,122.34) (42,124.97) (43,124.89) (44,121.59) (45,121.57)
  (46,122.08) (47,120.75) (48,121.59) (49,118.92) (50,118.42)
  (51,118.56) (52,122.63) (53,121.39) (54,120.73) (55,115.83)
  (56,112.38) (57,112.24) (58,109.63) (59,111.40) (60,107.03)
  (61,110.56) (62,113.25) (63,113.32) (64,109.76) (65,106.06)
  (66,106.24) (67,105.81) (68,109.75) (69,103.31) (70,100.26)
  (71,99.66) (72,100.36) (73,103.05) (74,102.76) (75,104.32)
  (76,107.80) (77,105.80) (78,105.28) (79,101.59) (80,101.60)
  (81,104.82) (82,105.61) (83,106.48) (84,104.81) (85,107.34)
  (86,105.56) (87,104.38) (88,102.06) (89,104.41) (90,107.31)
  (91,106.22) (92,106.69) (93,104.88) (94,104.43) (95,106.62)
  (96,104.61) (97,105.03) (98,106.88) (99,106.70) (100,109.27)
  (101,108.86) (102,109.41) (103,108.65) (104,108.06) (105,108.81)
  (106,110.50) (107,111.94) (108,112.08) (109,114.42) (110,115.36)
  (111,115.06) (112,115.29) (113,115.59) (114,116.54) (115,116.19)
  (116,117.17) (117,117.18) (118,118.02) (119,117.09) (120,117.87)
  (121,118.01) (122,117.99) (123,118.55) (124,119.00) (125,121.52)
  (126,121.74) (127,122.06) (128,119.10) (129,116.24) (130,117.85)
  (131,117.03) (132,118.28) (133,118.73) (134,115.83) (135,114.45)
  (136,111.28) (137,114.39) (138,114.47) (139,115.41) (140,119.23)
  (141,117.53) (142,116.67) (143,115.66) (144,118.79) (145,119.38)
  (146,116.18) (147,116.15) (148,117.29) (149,117.60) (150,117.96)
  (151,116.83) (152,118.40) (153,119.68) (154,119.26) (155,121.76)
  (156,123.24) (157,123.64) (158,124.84) (159,125.19) (160,125.25)
  (161,123.76) (162,124.53) (163,129.45) (164,131.25) (165,130.90)
  (166,130.00) (167,131.97) (168,131.84) (169,131.94) (170,130.02)
  (171,129.78) (172,131.13) (173,129.39) (174,132.30) (175,131.48)
  (176,130.07) (177,127.77) (178,129.57) (179,130.27) (180,131.01)
  (181,130.52) (182,129.36) (183,128.77) (184,132.53) (185,130.01)
  (186,131.29) (187,131.70) (188,129.74) (189,127.97) (190,128.09)
  (191,130.16) (192,128.11) (193,125.50) (194,128.89) (195,125.67)
  (196,125.23) (197,122.99) (198,122.73) (199,124.20) (200,126.50)
  (201,127.42) (202,129.08) (203,131.82) (204,131.33) (205,135.06)
  (206,136.74) (207,137.86) (208,139.81) (209,138.78) (210,141.22)
  (211,141.06) (212,141.34) (213,139.97) (214,137.28) (215,139.59)
  (216,137.93) (217,140.05) (218,140.71) (219,141.05) (220,140.33)
  (221,137.26) (222,137.58) (223,140.09) (224,139.54) (225,140.31)
  (226,141.29) (227,143.92) (228,143.75) (229,145.47) (230,145.33)
  (231,143.17) (232,145.70) (233,145.82) (234,144.87) (235,146.26)
  (236,148.76) (237,150.23) (238,147.12) (239,147.27) (240,145.78)
  (241,141.09) (242,139.51) (243,144.91) (244,149.71) (245,151.16)
  (246,149.57) (247,142.71) (248,144.97) (249,148.48) (250,148.96)
  (251,147.18) (252,146.90)
  };

  \draw[XBIa,line width=1pt] plot coordinates {  (0,100.00) 
  (1,100.49) (2,100.17) (3,99.02) (4,101.40) (5,104.03)
  (6,111.85) (7,117.98) (8,115.71) (9,116.38) (10,117.03)
  (11,119.42) (12,120.32) (13,122.90) (14,122.50) (15,122.47)
  (16,118.35) (17,113.38) (18,116.79) (19,115.40) (20,118.08)
  (21,115.37) (22,110.09) (23,111.64) (24,109.47) (25,108.76)
  (26,114.90) (27,118.16) (28,119.83) (29,120.11) (30,121.49)
  (31,119.40) (32,122.50) (33,120.48) (34,123.69) (35,126.26)
  (36,127.89) (37,130.95) (38,130.58) (39,131.07) (40,125.75)
  (41,125.43) (42,128.61) (43,128.15) (44,124.12) (45,123.54)
  (46,124.00) (47,122.75) (48,123.39) (49,121.06) (50,121.55)
  (51,120.78) (52,125.08) (53,124.55) (54,123.02) (55,117.87)
  (56,113.14) (57,112.58) (58,108.98) (59,109.97) (60,105.99)
  (61,109.66) (62,111.96) (63,111.14) (64,107.12) (65,102.72)
  (66,102.53) (67,102.84) (68,106.79) (69,99.89) (70,95.90)
  (71,94.26) (72,94.43) (73,97.16) (74,96.51) (75,98.88)
  (76,103.20) (77,100.34) (78,99.62) (79,95.72) (80,95.09)
  (81,98.71) (82,99.08) (83,100.33) (84,98.48) (85,100.32)
  (86,97.69) (87,96.36) (88,92.97) (89,95.58) (90,98.89)
  (91,97.94) (92,98.06) (93,96.67) (94,96.00) (95,98.09)
  (96,96.05) (97,95.96) (98,98.28) (99,99.11) (100,102.61)
  (101,102.28) (102,102.91) (103,101.83) (104,100.55) (105,100.77)
  (106,102.76) (107,104.92) (108,106.03) (109,112.57) (110,115.43)
  (111,114.54) (112,114.52) (113,114.08) (114,116.21) (115,116.45)
  (116,117.78) (117,118.00) (118,118.56) (119,117.83) (120,117.49)
  (121,117.78) (122,117.71) (123,118.59) (124,119.00) (125,121.08)
  (126,121.13) (127,121.24) (128,117.26) (129,112.84) (130,113.85)
  (131,112.56) (132,113.66) (133,113.94) (134,109.55) (135,107.93)
  (136,104.03) (137,107.48) (138,107.92) (139,108.67) (140,116.12)
  (141,114.82) (142,113.22) (143,111.39) (144,115.05) (145,116.71)
  (146,112.98) (147,112.09) (148,113.32) (149,114.12) (150,114.41)
  (151,112.79) (152,115.35) (153,117.34) (154,116.20) (155,118.77)
  (156,119.93) (157,120.05) (158,121.32) (159,120.95) (160,119.83)
  (161,117.59) (162,118.72) (163,122.57) (164,125.87) (165,125.32)
  (166,123.52) (167,125.15) (168,124.10) (169,124.64) (170,122.88)
  (171,121.96) (172,123.60) (173,121.42) (174,125.39) (175,125.75)
  (176,124.06) (177,121.25) (178,121.91) (179,123.48) (180,123.59)
  (181,122.74) (182,120.42) (183,119.63) (184,123.61) (185,121.34)
  (186,123.04) (187,123.91) (188,120.95) (189,119.37) (190,120.84)
  (191,122.78) (192,120.19) (193,117.31) (194,119.95) (195,116.34)
  (196,113.52) (197,112.79) (198,114.05) (199,118.04) (200,121.43)
  (201,121.04) (202,123.25) (203,124.58) (204,123.51) (205,128.19)
  (206,129.68) (207,129.67) (208,133.00) (209,131.54) (210,134.45)
  (211,132.91) (212,133.02) (213,131.89) (214,128.44) (215,131.06)
  (216,130.06) (217,133.50) (218,133.62) (219,134.69) (220,132.87)
  (221,130.22) (222,130.58) (223,132.90) (224,131.80) (225,133.82)
  (226,134.97) (227,138.20) (228,137.80) (229,139.98) (230,139.44)
  (231,135.48) (232,139.05) (233,139.18) (234,138.12) (235,140.58)
  (236,141.42) (237,145.50) (238,142.38) (239,142.72) (240,142.48)
  (241,136.40) (242,135.61) (243,142.08) (244,146.63) (245,147.99)
  (246,146.89) (247,139.31) (248,141.99) (249,145.26) (250,145.79)
  (251,144.26) (252,144.97)
  };
  
  \draw[FBTa,line width=1pt] plot coordinates {  (0,100.00) 
  (1,100.43) (2,99.87) (3,99.54) (4,101.31) (5,102.80)
  (6,104.12) (7,106.25) (8,104.02) (9,108.51) (10,108.72)
  (11,110.15) (12,110.12) (13,112.61) (14,112.08) (15,112.34)
  (16,108.27) (17,105.93) (18,107.91) (19,107.30) (20,110.42)
  (21,108.79) (22,105.12) (23,105.85) (24,104.86) (25,104.48)
  (26,108.57) (27,110.75) (28,112.11) (29,112.58) (30,113.86)
  (31,112.92) (32,115.90) (33,114.14) (34,116.70) (35,118.34)
  (36,119.57) (37,126.41) (38,123.91) (39,123.39) (40,120.61)
  (41,121.31) (42,123.75) (43,123.81) (44,120.74) (45,120.92)
  (46,121.44) (47,120.08) (48,120.99) (49,118.21) (50,117.38)
  (51,117.82) (52,121.81) (53,120.34) (54,119.96) (55,115.15)
  (56,112.13) (57,112.13) (58,109.84) (59,111.88) (60,107.37)
  (61,110.86) (62,113.67) (63,114.05) (64,110.64) (65,107.17)
  (66,107.47) (67,106.79) (68,110.74) (69,104.45) (70,101.72)
  (71,101.46) (72,102.34) (73,105.02) (74,104.84) (75,106.13)
  (76,109.34) (77,107.62) (78,107.17) (79,103.54) (80,103.77)
  (81,106.85) (82,107.79) (83,108.53) (84,106.92) (85,109.69)
  (86,108.18) (87,107.05) (88,105.09) (89,107.36) (90,110.12)
  (91,108.98) (92,109.57) (93,107.62) (94,107.24) (95,109.47)
  (96,107.46) (97,108.05) (98,109.74) (99,109.24) (100,111.49)
  (101,111.06) (102,111.58) (103,110.93) (104,110.57) (105,111.49)
  (106,113.08) (107,114.28) (108,114.09) (109,115.03) (110,115.34)
  (111,115.24) (112,115.54) (113,116.09) (114,116.65) (115,116.10)
  (116,116.97) (117,116.91) (118,117.84) (119,116.84) (120,118.00)
  (121,118.08) (122,118.08) (123,118.53) (124,119.01) (125,121.67)
  (126,121.94) (127,122.33) (128,119.72) (129,117.38) (130,119.18)
  (131,118.52) (132,119.82) (133,120.32) (134,117.92) (135,116.62)
  (136,113.70) (137,116.70) (138,116.65) (139,117.66) (140,120.27)
  (141,118.43) (142,117.82) (143,117.09) (144,120.03) (145,120.27)
  (146,117.25) (147,117.50) (148,118.62) (149,118.76) (150,119.14)
  (151,118.17) (152,119.41) (153,120.45) (154,120.28) (155,122.75)
  (156,124.34) (157,124.83) (158,126.02) (159,126.60) (160,127.06)
  (161,125.82) (162,126.47) (163,131.74) (164,133.04) (165,132.75)
  (166,132.16) (167,134.24) (168,134.42) (169,134.37) (170,132.39)
  (171,132.38) (172,133.64) (173,132.05) (174,134.60) (175,133.39)
  (176,132.08) (177,129.94) (178,132.12) (179,132.54) (180,133.48)
  (181,133.12) (182,132.34) (183,131.82) (184,135.50) (185,132.90)
  (186,134.04) (187,134.30) (188,132.67) (189,130.83) (190,130.50)
  (191,132.62) (192,130.75) (193,128.23) (194,131.87) (195,128.78)
  (196,129.13) (197,126.39) (198,125.63) (199,126.25) (200,128.19)
  (201,129.55) (202,131.02) (203,134.23) (204,133.94) (205,137.35)
  (206,139.09) (207,140.59) (208,142.08) (209,141.20) (210,143.48)
  (211,143.77) (212,144.12) (213,142.66) (214,140.23) (215,142.44)
  (216,140.55) (217,142.24) (218,143.08) (219,143.18) (220,142.82)
  (221,139.61) (222,139.91) (223,142.48) (224,142.12) (225,142.47)
  (226,143.39) (227,145.82) (228,145.74) (229,147.30) (230,147.30)
  (231,145.74) (232,147.92) (233,148.03) (234,147.12) (235,148.15)
  (236,151.21) (237,151.81) (238,148.70) (239,148.79) (240,146.88)
  (241,142.66) (242,140.81) (243,145.85) (244,150.74) (245,152.21)
  (246,150.46) (247,143.84) (248,145.97) (249,149.55) (250,150.02)
  (251,148.15) (252,147.54)
  };
\end{tikzpicture}
\end{figure}
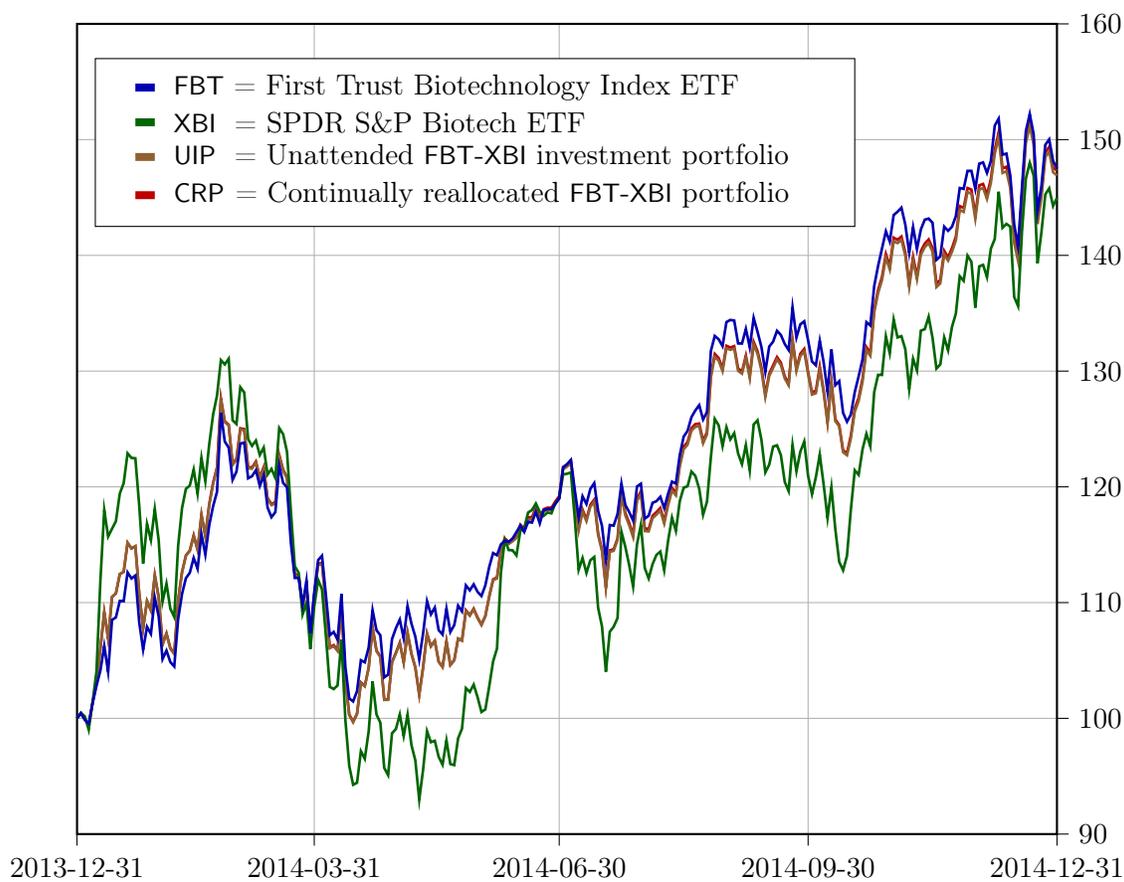

Figure \ref{FXUCplot} shows the 2014 history of an unattended
investment portfolio, \ticker{UIP}, in two high-returning exchange
traded funds, \ticker{FBT} and \ticker{XBI}. The 2013-12-31 
closing composition
of \ticker{UIP} was\\[-1.5ex]
\begin{equation}\label{UIPportfolio_1}
  \ticker{UIP} = 75\%~\ticker{FBT} + 25\%~\ticker{XBI},
\end{equation}
but these closing proportions never reoccured in 2014. Indeed
\ticker{UIP} closed with more than 25\% in \ticker{XBI} (green higher
than blue) for most of the first quarter, whereas \ticker{XBI} was less
than 25\% of \ticker{UIP} (blue higher than green) for much of the
remaining three quarters. However Figure \ref{FXUCplot} does shows the
\emph{geometry} of the 3:1 proportions. On every vertical line, the
brown \ticker{UIP} point is exactly 3/4 of the way from the green
\ticker{XBI} point to the blue \ticker{FBT} point.

There were exactly 253 market days from 2013-12-31 to 2014-12-31
inclusive. Each of the four adjusted closing price graphs in Figure
\ref{FXUCplot} represents the changing value of \$100 invested at
2013-12-31 closing prices in the corresponding fund or portfolio
over the next 252 market days. Each graph corresponds to a point
$\mathbf{a}\in\mathds{R}^{253}$, and the three points,
$\mathbf{a}_{\sstick{FBT}},\mathbf{a}_{\sstick{XBI}},~\text{and}~
\mathbf{a}_{\sstick{UIP}}$ of \eqref{UIPportfolio_1}, are on the
line segment%
\vspace*{-1ex}
\begin{equation}\label{FBTXBI_line}
  \mathbf{a} = t\cdot\mathbf{a}_{\sstick{FBT}} +
  (1 - t)\cdot\mathbf{a}_{\sstick{XBI}}, \quad 0 \le t \le 1,
\end{equation}
in $\mathds{R}^{253}$---with $\mathbf{a}_{\sstick{UIP}}$
corresponding to $t = 0.75$. (Appendix \ref{adjclose_primer} describes
how normalized adjusted closing prices can be computed.)

\newpage
Portfolio \ticker{UIP} represents a completely passive, unattended
investment. Think of an investor as \emph{having} money in \ticker{FBT}
and \ticker{XBI} at the close of 2013. Suppose that \ticker{FBT}
represents exactly 75\% of his total investment at that time and
\ticker{XBI} 25\%. \ticker{UIP} then simply tracks each \$100 of his
investment through 2014. The investor does absolutely nothing, and all
dividends from the two funds are automatically reinvested.

However the continually reallocated portfolio \ticker{CRP}, mostly
hidden by \ticker{UIP}, is an entirely different matter. Here the
investor decides, a priori, that 75\% FBT and 25\% XBI are the right
proportions for his investment. Accordingly, before each market day of
2014, he reinvests his money so as to start the day with exactly these
proportions in the two funds.\\
Algorithm \ref{CRP} computes the growth of \$100 under this scenario.

\begin{algorithm}[H]
  \centering
  \caption{\label{CRP}%
    To compute the 2014 \ticker{CRP} adjusted closing price vector~~~~~
  }
\vspace*{-1.5ex}%
\begin{tikzpicture}[scale=1]
\node[right] at (0,0){
\parbox{12.8cm}{$ 
\begin{array}{|l|l}
  \multicolumn{2}{l}{a_{0,\sstick{CRP}}=100;
    ~~\textcolor{algcomment}{\text{\% invest \$100 in \ticker{CRP} at the 2013-12-31 close}}}\\
  \multicolumn{2}{l}{\text{for}~ i=1,\ldots,252
    ~~\textcolor{algcomment}{\text{\% for each of the 252 market days in 2014 set}}}\\
  \multicolumn{1}{p{1ex}}{} &
  \multicolumn{1}{l}{r_{\sstick{F}}=a_{i,\sstick{FBT}}/a_{i-1,\sstick{FBT}}-1;
    ~~\textcolor{algcomment}{\text{\% = return of \ticker{FBT} on market day}~ i}}\\
  \multicolumn{1}{p{1ex}}{} &
  \multicolumn{1}{l}{r_{\sstick{X}}=a_{i,\sstick{XBI}}/a_{i-1,\sstick{XBI}}-1;
    ~~\textcolor{algcomment}{\text{\% = return of \ticker{XBI} on market day}~ i}}\\
  \multicolumn{1}{p{1ex}}{} &
  \multicolumn{1}{l}{r_{\sstick{R}}=0.75\times r_{\sstick{F}}+0.25\times r_{\sstick{X}};
    ~~\textcolor{algcomment}{\text{\% = return of \ticker{CRP} on market day}~ i}}\\
  \multicolumn{1}{p{1ex}}{} &
  \multicolumn{1}{l}{a_{i,\sstick{CRP}}=a_{i-1,\sstick{CRP}}\times(1+r_{\sstick{R}});
    ~~\textcolor{algcomment}{\text{\% = \ticker{CRP} value at the close of market day}~ i}}
\end{array}$}};
\draw (0.35,0.48) -- (0.35,-1.4) -- (0.7,-1.4);
\end{tikzpicture}
\end{algorithm}

\begin{note*}
  The \textcolor{algcomment}{\%} signs above begin a
  \textcolor{algcomment}{comment}.
\end{note*}

\begin{minipage}{9.2cm}
It is difficult to make out the red \ticker{CRP} graph from the brown
\ticker{UIP} graph in Figure \ref{FXUCplot}. These graphs are very close,
and the \ticker{UIP} graph is drawn over the \ticker{CRP} graph, hiding it
from view for the most part. It is only toward the end of 2014 that one
can really make out the differences.\\

Figure \ref{FXUCplot2} is a blow-up of December 2014. The differences in
the graphs are now quite visible. Here we see that the red \ticker{CRP}
graph is slightly higher than the brown \ticker{UIP} graph throughout
December---and clearly higher on December 31. It follows that
\ticker{CRP} returned more than \ticker{UIP} over 2014.\\

Note that the 2014-12-31 value of $\mathbf{a}_{\ticker{UIP}}$
\emph{must} be \[ 146.90 = 0.75 \times 147.54 + 0.25 \times 144.97 \] by
\eqref{convex_comb} and the terminal values of \ticker{FBT} and
\ticker{XBI} shown on Figure \ref{FXUCplot2}. It follows that
\ticker{UIP} had a total return of 46.90\% over 2014. In fact
\ticker{CRP} returned 47.32\% over 2014, more than \ticker{UIP} and just
slightly less than \ticker{FBT}.
\end{minipage}
\hfill
\begin{minipage}{5.3cm}
\begin{figure}[H]
  \centering
  \caption{\label{FXUCplot2}December~~~~~}
  \vspace*{-2ex}%
\begin{tikzpicture}[scale=3.8,xscale=0.03,yscale=0.12]

  \draw[grid] (231,140) node[black,left]{140} -- (252, 140);
  \draw[grid] (231,145) node[black,left]{145} -- (252, 145);
  \draw[grid] (231,150) node[black,left]{150} -- (252, 150);
  \node[left] at (231,135) {135};
  \draw[thick] (231,153) -- (231,153.3) node[left=2,above]{Dec 1};
  \draw[thick] (252,153) -- (252,153.3) node[left=3,above]{Dec 31};
  
  \draw[CRPa,line width=1.25pt] plot coordinates { (231,143.49)  
  (232,146.05) (233,146.17) (234,145.22) (235,146.62) (236,149.11)
  (237,150.63) (238,147.51) (239,147.66) (240,146.18) (241,141.47)
  (242,139.89) (243,145.31) (244,150.13) (245,151.58) (246,149.99)
  (247,143.11) (248,145.38) (249,148.90) (250,149.38) (251,147.59)
  (252,147.32)
  };
  
  \draw[UIPa,line width=1.25pt] plot coordinates { (231,143.17)  
  (232,145.70) (233,145.82) (234,144.87) (235,146.26) (236,148.76)
  (237,150.23) (238,147.12) (239,147.27) (240,145.78) (241,141.09)
  (242,139.51) (243,144.91) (244,149.71) (245,151.16) (246,149.57)
  (247,142.71) (248,144.97) (249,148.48) (250,148.96) (251,147.18)
  (252,146.90)
  };
  
  \draw[XBIa,line width=1.25pt] plot coordinates { (231,135.48)  
  (232,139.05) (233,139.18) (234,138.12) (235,140.58) (236,141.42)
  (237,145.50) (238,142.38) (239,142.72) (240,142.48) (241,136.40)
  (242,135.61) (243,142.08) (244,146.63) (245,147.99) (246,146.89)
  (247,139.31) (248,141.99) (249,145.26) (250,145.79) (251,144.26)
  (252,144.97)}
  -- (253.5,144.97) node[XBIa,right]{144.97};
  \node[XBIa,right] at (255.7,144) {\ticker{XBI}};
  
  \draw[FBTa,line width=1.25pt] plot coordinates { (231,145.74)  
  (232,147.92) (233,148.03) (234,147.12) (235,148.15) (236,151.21)
  (237,151.81) (238,148.70) (239,148.79) (240,146.88) (241,142.66)
  (242,140.81) (243,145.85) (244,150.74) (245,152.21) (246,150.46)
  (247,143.84) (248,145.97) (249,149.55) (250,150.02) (251,148.15)
  (252,147.54)}
  -- (253.5,147.54) node[FBTa,right]{147.54};
  \node[FBTa,right] at (255.4,148.42) {\ticker{FBT}};
  
  \draw[thick] (231,135) rectangle (252,153);
  \node at (241,134) {2014};
\end{tikzpicture}
\end{figure}
\vspace*{0.2ex}%
\end{minipage}

\newpage

Here are the adjusted closing prices of the four funds over the month
of December 2014.

\begin{table}[H]
  \centering
  \caption{\label{decprices}%
  The December adjusted closing prices
  }
\begin{tabular}{|c|rrrr|}\hline\rule{0mm}{4mm}
  date & \smtick{FBT}~~ & \smtick{XBI}~~ & \smtick{UIP}~~
  & \smtick{CRP}~~ \\\hline\rule{0mm}{3.8mm}%
  2014-12-01 & 145.736 & 135.477 & 143.171 & 143.490 \\
  2014-12-02 & 147.918 & 139.052 & 145.702 & 146.048 \\
  2014-12-03 & 148.034 & 139.184 & 145.821 & 146.169 \\
  2014-12-04 & 147.123 & 138.121 & 144.873 & 145.215 \\
  2014-12-05 & 148.150 & 140.580 & 146.258 & 146.622 \\
  2014-12-08 & 151.214 & 141.417 & 148.765 & 149.114 \\
  2014-12-09 & 151.807 & 145.504 & 150.231 & 150.630 \\
  2014-12-10 & 148.699 & 142.379 & 147.119 & 147.508 \\
  2014-12-11 & 148.786 & 142.720 & 147.269 & 147.661 \\
  2014-12-12 & 146.878 & 142.480 & 145.778 & 146.179 \\
  2014-12-15 & 142.657 & 136.400 & 141.093 & 141.469 \\
  2014-12-16 & 140.807 & 135.609 & 139.507 & 139.888 \\
  2014-12-17 & 145.851 & 142.076 & 144.907 & 145.314 \\
  2014-12-18 & 150.737 & 146.628 & 149.710 & 150.129 \\
  2014-12-19 & 152.212 & 147.990 & 151.156 & 151.579 \\
  2014-12-22 & 150.463 & 146.886 & 149.569 & 149.990 \\
  2014-12-23 & 143.842 & 139.313 & 142.710 & 143.107 \\
  2014-12-24 & 145.968 & 141.988 & 144.973 & 145.380 \\
  2014-12-26 & 149.554 & 145.261 & 148.481 & 148.897 \\
  2014-12-29 & 150.017 & 145.790 & 148.960 & 149.378 \\
  2014-12-30 & 148.151 & 144.258 & 147.178 & 147.592 \\
  2014-12-31 & 147.544 & 144.973 & 146.901 & 147.321 \\
  \hline
\end{tabular}
\end{table}

The geometric equation \eqref{UIPportfolio_1} holds on every line of
Table \ref{decprices}, but the proportions of \ticker{FBT} and
\ticker{XBI} in \ticker{UIP},\\[-2.5ex]
\begin{equation}\label{UIP_prportions}
  p_{\sstick{FBT}} = 0.75 \times \ticker{FBT} / \ticker{UIP}
  \text{~~and~~}
  p_{\sstick{XBI}} = 0.25 \times \ticker{XBI} / \ticker{UIP},
\end{equation}
are different on every line. The 2013-12-31, 3:1 proportions
come closest to being realized on the 2014-12-31 line of
Table \ref{decprices}, where \ticker{UIP} closes at
75.33\%\,\ticker{FBT}\,:\,24.67\%\,\ticker{XBI}.

As for daily returns,\\[-2ex]
\begin{equation}\label{daily_returns}
  r_i = a_i/a_{i-1} - 1 \quad(i = 1,\ldots,252),
\end{equation}
the 2014 return vector equation,\\[-2ex]
\begin{equation}\label{CRP_dailyreturn}
  \mathbf{r}_{\sstick{CRP}} = 0.75~ \mathbf{r}_{\sstick{FBT}}
  + 0.25~  \mathbf{r}_{\sstick{XBI}},
\end{equation}
is valid in $\mathds{R}^{252}$ due to continual reallocation. This
insures that the corresponding \emph{mean} returns satisfy\\[-4ex]
\begin{equation}\label{expected_mean_return}
  e_{\sstick{CRP}} = 0.75~ e_{\sstick{FBT}} + 0.25~ e_{\sstick{XBI}},
\end{equation}
as required by ``The Standard Mean-Variance Portfolio
Selection Model'' of  \cite[pp. 3-4]{Markowitz:1987wd}.

\vspace*{2ex}
The \verb!ancillary! folder that accompanies this article includes three
files, \verb!FXUZ7.csv!,\\\verb!matlab/FXUC2014.mat!, and
\verb!matlab/FXZC2014.mat!, that contain the normalized adjusted closing
prices used for this article.

\newpage
\section{How not to do mean-variance analysis}
\label{how-not-to}

The \href{https://www.mathworks.com/products/finance.html}
{$\text{MathWorks}^{\text{\tiny{\textregistered}}}$} Financial Toolbox
with the MATLAB programming language is perhaps the most popular
resource for doing mean-variance analysis. We have computed our
mean-variance tables using MATLAB scripts
in the \verb!matlab! subdirectory of the \verb!ancillary!
folder that accompanies this article.

Our MATLAB script \verb!hn2mv1a.m! illustrates the problem with
mean-variance analysis as it is usually practiced. The script
begins with the line\\[0.5ex]
\hspace*{3cm}%
  \lstinline!load FXUC2014.mat;  % A  dates  fundsA  legendA!\\[0.5ex]
which loads the $253\times4$ matrix of adjusted closing prices, $A =
\left[~\mathbf{a}_{\sstick{FBT}},~ \mathbf{a}_{\sstick{XBI}},~
\mathbf{a}_{\sstick{UIP}},~ \mathbf{a}_{\sstick{CRP}}\,\right],$\\
displayed in Figure \ref{FXUCplot}. This matrix has rank 3 rather than
4, since $\mathbf{a}_{\sstick{UIP}}$ is a linear combination of
$\mathbf{a}_{\sstick{FBT}}$ and $\mathbf{a}_{\sstick{XBI}}$
\eqref{convex_comb}.

Next we remove $\mathbf{a}_{\sstick{CRP}}$ from $A$ and append the
columns\\\hspace*{2cm}%
$ \mathbf{a}_{\sstick{UIP2}} = 0.50 \cdot \mathbf{a}_{\sstick{FBT}}
  + 0.50 \cdot \mathbf{a}_{\sstick{XBI}}~~\text{and}~~
  \mathbf{a}_{\sstick{UIP3}} = 0.25 \cdot \mathbf{a}_{\sstick{FBT}}
  + 0.75 \cdot \mathbf{a}_{\sstick{XBI}}$\\[0.5ex]
so that ~$%
A = \left[~\mathbf{a}_{\sstick{FBT}},~ \mathbf{a}_{\sstick{XBI}},~
\mathbf{a}_{\sstick{UIP}},~ \mathbf{a}_{\sstick{UIP2}},~
\mathbf{a}_{\sstick{UIP3}}\,\right]$~
becomes a $253\times5$ matrix of rank 2. Geometrically, $A$ describes 5
points on a line in $\mathds{R}^{253}$, and $\verb!rank!(A)=2$ since the
line does not pass through the origin.

The \verb!hn2mv1a.m! script continues with the lines\\
\hspace*{\fill}%
\begin{minipage}{14cm}
\begin{lstlisting}
%% get asset moments from adjusted closing prices
ptf = Portfolio;
ptf = estimateAssetMoments(ptf, A, 'dataformat', 'prices');
[mn, cv] = getAssetMoments(ptf);
\end{lstlisting}
\end{minipage}
Here \lstinline!mn! and \lstinline!cv! are the $5\times1$ and
$5\times5$ mean daily return and covariance of
daily return matrices corresponding to the $252\times5$ daily return
matrix\\\hspace*{2cm}%
$R = \left[~\mathbf{r}_{\sstick{FBT}},~ \mathbf{r}_{\sstick{XBI}},~
\mathbf{r}_{\sstick{UIP}},~ \mathbf{r}_{\sstick{UIP2}},~
\mathbf{r}_{\sstick{UIP2}}\,\right]$\\[0.5ex]
derived from $A$ via \eqref{daily_returns}. The annualized results are
shown in Table \ref{hn2mv1_table}.

\begin{table}[H]
  \centering
  \caption{\label{hn2mv1_table}%
    Annualized results of the MATLAB computations}
  \vspace*{-1ex}
\begin{tabular}{|l|rrrrr|}\cline{2-6}
  \multicolumn{1}{c|}{\rule{0mm}{3.8mm}}
  & \smtick{FBT}~ & \smtick{XBI}~~ & \smtick{UIP}~~
  & \smtick{UIP2}~~ & \smtick{UIP3}~~ \\\hline\rule{0mm}{3.8mm}%
  $E$ & 0.4245 & 0.4324 & 0.4235 & 0.4244 & 0.4274 \\
  $\lgsig$ & 0.2656 & 0.3491 & 0.2777 & 0.2963 & 0.3203 \\\hline
  \multicolumn{6}{c}{\hspace*{5ex}covariance $V$\rule{0mm}{4mm}}\\
  \hline\rule{0mm}{3.8mm}%
  \smtick{FBT} & 0.0705 & 0.0804 & 0.0729 & 0.0753 & 0.0778 \\
  \smtick{XBI} & 0.0804 & 0.1219 & 0.0905 & 0.1008 & 0.1112 \\
  \smtick{UIP} & 0.0729 & 0.0905 & 0.0771 & 0.0815 & 0.0859 \\
  \smtick{UIP2} & 0.0753 & 0.1008 & 0.0815 & 0.0878 & 0.0942 \\
  \smtick{UIP3} & 0.0778 & 0.1112 & 0.0859 & 0.0942 & 0.1026 \\\hline
\end{tabular}
\end{table}

One problem with Table \ref{hn2mv1_table} is immediately obvious. How can the
mean returns of the long portfolios \ticker{UIP} and \ticker{UIP2} be
less than the mean return of either component fund? A dimensional problem
is less obvious but just as troubling. We start with an adjusted
closing price history, $A$, which corresponds to five points on a line
segment (a one simplex) in $\mathds{R}^{253}$;~ but, to do mean-variance
analysis on $A$, we must jump to the four simplex in $\mathds{R}^{252}$
generated by the five linearly independent columns of the daily return matrix
$R$ ($\verb!rank!(R)=\verb!rank!(V)=5$). It simply doesn't make sense to us!

\newpage

\begin{minipage}{7.1cm}
\begin{minipage}{6.5cm}
This is the $(e,\sigma)$-picture of what is happening. The pink region
is the image of the 4-simplex in $\mathds{R}^{252}$ generated by the
five columns of $R$.
(The coordinates in the picture are percentages.)
\end{minipage}\\
\begin{minipage}{7.1cm}
\begin{figure}[H]
  \centering
  \caption{\label{reallocated_esig}Obtainable $(e,\sigma)$\\[-1ex]}
\hspace*{-1.5cm}%
\begin{tikzpicture}[scale=0.56,>={angle 60}]
  \setlength{\pointradius}{0.15cm}
  \setlength{\smallradius}{0.7\pointradius}
  
  \newcommand{\xL}{0}  
  \newcommand{\xH}{10}  
  \newcommand{\dx}{3.33333}  
  \newcommand{\yL}{25}      
  \newcommand{\yH}{37}      
  \newcommand{\dy}{5}
  
  \coordinate (FBT)  at (3.014,26.61);  
  \coordinate (UIP)  at (2.324,27.77);  
  \coordinate (UIP2) at (2.943,29.63);
  \coordinate (UIP3) at (4.915,32.03);
  \coordinate (XBI)  at (8.300,34.91);  
  \coordinate (CRP)  at (4.335,27.83);  
  
  \draw (\xL,\yL) rectangle (\xH,\yH);
  \draw (\xL,\yL) -- (\xL,\yL-0.2) node[below]{42.0};
  \draw (\xL+\dx,\yL) -- (\xL+\dx,\yL-0.2) node[below]{42.5};
  \draw (\xL+2*\dx,\yL) -- (\xL+2*\dx,\yL-0.2) node[below]{43.0};
  \draw (\xH,\yL) -- (\xH,\yL-0.2) node[below]{43.5};
  \draw (\xL,\yL) -- (\xL-0.2,\yL) node[left]{25};
  \draw (\xL,\yL+5) -- (\xL-0.2,\yL+5) node[left]{30};
  \draw (\xL,\yL+10) -- (\xL-0.2,\yL+10) node[left]{35};
  
  \draw[thick,->] (\xL-0.2,\yL) -- (\xH+1,\yL) node[right]{$e$};
  \draw[thick,->] (\xL,\yL-0.2) -- (\xL,\yH+0.8) node[above]{$\sigma$};
  
  \filldraw[fill=CRi,draw=CR] plot[smooth] coordinates {
  (3.014,26.56)  
  (3.542,26.98) (4.071,27.52) (4.600,28.16) (5.128,28.89) (5.657,29.72)
  (6.185,30.62) (6.714,31.60) (7.243,32.64) (7.771,33.75) (8.300,34.91)
  } -- plot[smooth] coordinates {
  (8.300,34.91)  
  (7.961,34.61) (7.623,34.31) (7.284,34.01) (6.946,33.72) (6.608,33.43)
  (6.269,33.14) (5.931,32.86) (5.592,32.58) (5.254,32.30) (4.915,32.03)
  } -- plot[smooth] coordinates {
  (4.915,32.03)  
  (4.718,31.77) (4.521,31.52) (4.323,31.26) (4.126,31.02) (3.929,30.77)
  (3.732,30.53) (3.535,30.30) (3.337,30.07) (3.140,29.84) (2.943,29.63)
  } -- plot[smooth] coordinates {
  (2.943,29.63)  
  (2.881,29.42) (2.819,29.21) (2.757,29.01) (2.696,28.82) (2.634,28.63)
  (2.572,28.45) (2.510,28.27) (2.448,28.10) (2.386,27.93) (2.324,27.77)
  } -- plot[smooth] coordinates {
  (2.324,27.77)  
  (2.393,27.62) (2.462,27.48) (2.531,27.34) (2.600,27.21) (2.669,27.08)
  (2.738,26.96) (2.807,26.85) (2.876,26.75) (2.945,26.65) (3.014,26.56)
  };
  
  \draw[CR,thick] plot[smooth] coordinates {
  (3.014,26.56)  
  (3.542,26.98) (4.071,27.52) (4.600,28.16) (5.128,28.89) (5.657,29.72)
  (6.185,30.62) (6.714,31.60) (7.243,32.64) (7.771,33.75) (8.300,34.91)
  };
  
  \draw[XF,thick] plot[smooth] coordinates {
  (3.014,26.56)  
  (2.887,26.65) (2.773,26.74) (2.672,26.85) (2.584,26.96) (2.508,27.08)
  (2.445,27.20) (2.396,27.34) (2.359,27.47) (2.335,27.62) (2.324,27.77)
  (2.327,27.93) (2.342,28.09) (2.371,28.26) (2.413,28.44) (2.468,28.62)
  (2.536,28.81) (2.618,29.01) (2.713,29.21) (2.821,29.41) (2.943,29.63)
  (3.078,29.84) (3.227,30.07) (3.390,30.29) (3.566,30.53) (3.756,30.77)
  (3.960,31.01) (4.178,31.26) (4.410,31.51) (4.655,31.77) (4.915,32.03)
  (5.189,32.30) (5.477,32.57) (5.779,32.85) (6.096,33.13) (6.427,33.42)
  (6.772,33.71) (7.132,34.00) (7.507,34.30) (7.896,34.61) (8.300,34.91)
  };
  
  \draw[grid] (\dx,\yL) -- (\dx,\yH);
  \draw[grid] (2*\dx,\yL) -- (2*\dx,\yH);
  \draw[grid] (\xL,27.5) -- (\xH,27.5);
  \draw[grid] (\xL,30) -- (\xH,30);
  \draw[grid] (\xL,32.5) -- (\xH,32.5);
  \draw[grid] (\xL,35) -- (\xH,35);
  
  \node[rotate=53] at (5.1,30.75) {\parbox{3cm}{continually\\reallocated}};
  \node[rotate=42.6] at (4.8,32.7) {unattended path};
  \node[rotate=64] at (7,31) {efficient frontier};
  
  \filldraw[fill=FBTv,thick] (FBT) circle[radius=\pointradius]
    node[below=2]{\smtick{FBT}};
  \filldraw[fill=XBIv,thick] (XBI) circle[radius=\pointradius]
    node[right=1,above=2]{\smtick{XBI}};
  \filldraw[fill=UIPv,thick] (UIP) circle[radius=\pointradius]
    node[left=1]{\smtick{UIP}};
  \filldraw[fill=UIPv,thick] (UIP2) circle[radius=0.8\pointradius]
    node[left=1]{\smtick{UIP2}};
  \filldraw[fill=UIPv,thick] (UIP3) circle[radius=0.8\pointradius]
    node[below=1,right=1]{\smtick{UIP3}};
  \filldraw[fill=CRPv,thick] (CRP) circle[radius=\pointradius]
    node[right=9,below=1]{\smtick{CRP}};
  \fill[black] (5.657,29.72) circle (\smallradius);
  \fill[black] (6.979,32.11) circle (\smallradius);
  
\begin{scope}[scale=2.65,xshift=-1.4cm,yshift=-20.2cm]
  \coordinate (FBT)  at (3.014,26.61);  
  \coordinate (UIP)  at (2.324,27.77);  
  \coordinate (UIP2) at (2.943,29.63);
  \coordinate (UIP3) at (4.915,32.03);
  \coordinate (XBI)  at (8.300,34.91);  
  \coordinate (CRP)  at (4.335,27.83);  
  
  \begin{scope}
    \clip (\xL,\yL) rectangle (FBT);
  \end{scope}

  \draw[clip] (3.29, 27.6) circle[radius=1.6cm];
  
  \filldraw[fill=CRi,draw=CR] plot[smooth] coordinates {
  (3.014,26.56)  
  (3.542,26.98) (4.071,27.52) (4.600,28.16) (5.128,28.89) (5.657,29.72)
  (6.185,30.62) (6.714,31.60) (7.243,32.64) (7.771,33.75) (8.300,34.91)
  } -- plot[smooth] coordinates {
  (8.300,34.91)  
  (7.961,34.61) (7.623,34.31) (7.284,34.01) (6.946,33.72) (6.608,33.43)
  (6.269,33.14) (5.931,32.86) (5.592,32.58) (5.254,32.30) (4.915,32.03)
  } -- plot[smooth] coordinates {
  (4.915,32.03)  
  (4.718,31.77) (4.521,31.52) (4.323,31.26) (4.126,31.02) (3.929,30.77)
  (3.732,30.53) (3.535,30.30) (3.337,30.07) (3.140,29.84) (2.943,29.63)
  } -- plot[smooth] coordinates {
  (2.943,29.63)  
  (2.881,29.42) (2.819,29.21) (2.757,29.01) (2.696,28.82) (2.634,28.63)
  (2.572,28.45) (2.510,28.27) (2.448,28.10) (2.386,27.93) (2.324,27.77)
  } -- plot[smooth] coordinates {
  (2.324,27.77)  
  (2.393,27.62) (2.462,27.48) (2.531,27.34) (2.600,27.21) (2.669,27.08)
  (2.738,26.96) (2.807,26.85) (2.876,26.75) (2.945,26.65) (3.014,26.56)
  };
  
  \draw[XF,very thick] plot[smooth] coordinates {
  (3.014,26.56)  
  (2.887,26.65) (2.773,26.74) (2.672,26.85) (2.584,26.96) (2.508,27.08)
  (2.445,27.20) (2.396,27.34) (2.359,27.47) (2.335,27.62) (2.324,27.77)
  (2.327,27.93) (2.342,28.09) (2.371,28.26) (2.413,28.44) (2.468,28.62)
  (2.536,28.81) (2.618,29.01) (2.713,29.21) (2.821,29.41) (2.943,29.63)
  (3.078,29.84) (3.227,30.07) (3.390,30.29) (3.566,30.53) (3.756,30.77)
  (3.960,31.01) (4.178,31.26) (4.410,31.51) (4.655,31.77) (4.915,32.03)
  (5.189,32.30) (5.477,32.57) (5.779,32.85) (6.096,33.13) (6.427,33.42)
  (6.772,33.71) (7.132,34.00) (7.507,34.30) (7.896,34.61) (8.300,34.91)
  };
  
  \draw[CR,very thick] plot[smooth] coordinates {
  (3.014,26.56)  
  (3.542,26.98) (4.071,27.52) (4.600,28.16) (5.128,28.89) (5.657,29.72)
  (6.185,30.62) (6.714,31.60) (7.243,32.64) (7.771,33.75) (8.300,34.91)
  };
  
  \draw[grid] (\dx,\yL) -- (\dx,\yH);
  \draw[grid] (\xL,27.5) -- (\xH,27.5);
  
  \node at (3.80,28) {\parbox{3cm}{continually\\reallocated\\portfolios}};
  
  \filldraw[fill=FBTv,thick] (FBT) circle[radius=0.065cm]
    node[below=2]{\smtick{FBT}};
  \filldraw[fill=UIPv,thick] (UIP) circle[radius=0.065cm]
    node[left=1]{\smtick{UIP}};
  \filldraw[fill=CRPv,thick] (CRP) circle[radius=0.065cm]
    node[right=9,below=1]{\smtick{CRP}};
\end{scope}
\end{tikzpicture}
\end{figure}
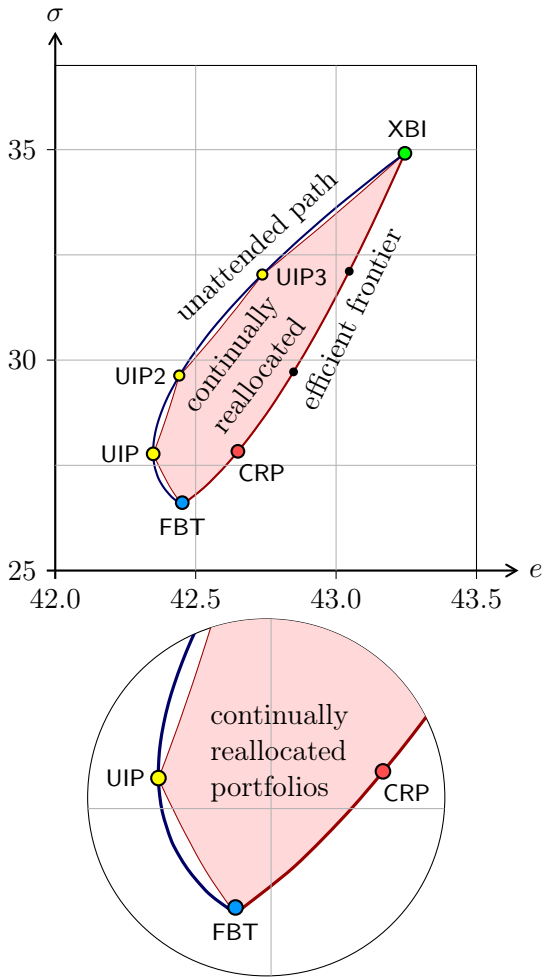
\end{minipage}
\end{minipage}
\hfill%
\begin{minipage}{7.5cm}
\vspace*{-1.8cm}%
Figure \ref{reallocated_esig} is a graphic representation of
Table \ref{hn2mv1_table}. The red, continually reallocated region
represents all obtainable $(e_p,\sigma_p)$\,:\\
all $(e_p,\sigma_p)$ such that\\[-4ex]
\begin{gather}\label{esigv}
  e_p = E \mathbf{p},~~ \sigma_p = \sqrt{\vphantom{b} v_p},~~
  v_p = \mathbf{p}^T V \mathbf{p},\\[-1ex]
\intertext{with the $E$ and $V$ from Table \ref{hn2mv1_table}, and
  \vspace*{-1ex}}
\label{longproportions}
  \bm{0}\le\mathbf{p}\le\bm{1},~~ \sum\mathbf{p} = 1,
  ~~ \mathbf{p}\in\mathds{R}^5.
\end{gather}

The following five portfolios $\mathbf{p}$ are equally $e$-spaced
on the efficient frontier. They were computed with the line\\
\hspace*{1.4cm}\lstinline!P = estimateFrontier(ptf, 5);!\\
in the MATLAB script \verb!hn2mv1a.m!.\\

\hspace*{2ex}%
$P =
\begin{blockarray}{rccccl}
  \smtick{FBT} & \smtick{CRP} & \smtick{???} & \smtick{???}
   & \smtick{XBI} & \\
\begin{block}{[r@{\hspace*{2.5ex}}rrrr]l}
   1 & 0.75 & 0.50 & 0.25 & 0\,~ & \smtick{FBT} \\
   0 & 0.25 & 0.50 & 0.75 & 1\,~ & \smtick{XBI} \\
   0 & 0~~ & 0~~ & 0~~ & 0\,~ & \smtick{UIP} \\
   0 & 0~~ & 0~~ & 0~~ & 0\,~ & \smtick{UIP2} \\
   0 & 0~~ & 0~~ & 0~~ & 0\,~ & \smtick{UIP3} \\
\end{block}
\end{blockarray}
$

Every portfolio in the continually reallocated region, other than
the generating funds \ticker{FBT}, \ticker{XBI}, \ticker{UIP},
\ticker{UIP2}, and \ticker{UIP3}, must be reallocated at the close each
market day of 2014 in order to retain its original \mbox{2013-12-31}
closing proportions.\\

On the other hand, the unattended path in Figure \ref{reallocated_esig}
shows the actual mean returns and standard deviations of return of
all unattended portfolios in \ticker{FBT} and \ticker{XBI} as computed
directly from their adjusted closing price vectors \eqref{FBTXBI_line}
via MATLAB Example \ref{matlab2} (script \verb!hn2mv2.m!) below.
\end{minipage}

\vspace*{-1.2cm}%
\begin{matlab}[H]
  \captionsetup{margin={-2.8ex,2.8ex}}
  \caption{\label{matlab2}%
    To compute the unattended portfolio path from \ticker{FBT} to \ticker{XBI} do
  }
  \vspace*{-0.6cm}
\begin{lstlisting}
T = 0$\,$:$\,$1/(nT - 1)$\,$:$\,$1;  % nT points partitioning [0, 1]
AT = A(:,$\,$1)$\,$*$\,$$\,$T + A(:,$\,$2)$\,$*$\,$(1 - T); %  nT unattended price vectors
RT = AT(2:253,$\,$:)$\,$./$\,$ AT(1:252,$\,$:) - 1; %  nT unattended return vectors
ET = 252$\,$*$\,$mean(RT);  % nT mean daily returns (annualized)
SigT = sqrt(252)$\,$*$\,$std(RT, 1);  % nT standard deviations of return
\end{lstlisting}
\end{matlab}

\vspace*{-3ex}%
We should note that the MATLAB lines
of \verb!hn2mv1a.m!,\\[1ex]
\hspace*{1cm}%
\begin{minipage}{12cm}
\lstinline!eCRP = 252 * mean(rCRP);                   %  eCRP    = 0.4265!\\
\lstinline!sigCRP = sqrt(252) * std(rCRP, 1);         %  sigCRP = 0.2783!
\end{minipage}\\[1ex]
produced the coordinates for the two \ticker{CRP} points of Figure
\ref{reallocated_esig}. The \lstinline!rCRP! in this code
corresponds to the $\mathbf{r}_{\sstick{CRP}}$ vector of
\eqref{CRP_dailyreturn} or the $\mathbf{r}_{\sstick{CRP}}$ from
$\mathbf{a}_{\sstick{CRP}}$ via \eqref{daily_returns}. They are the same.

\newpage
\section{The mean periodic return problem}\label{return_problem}

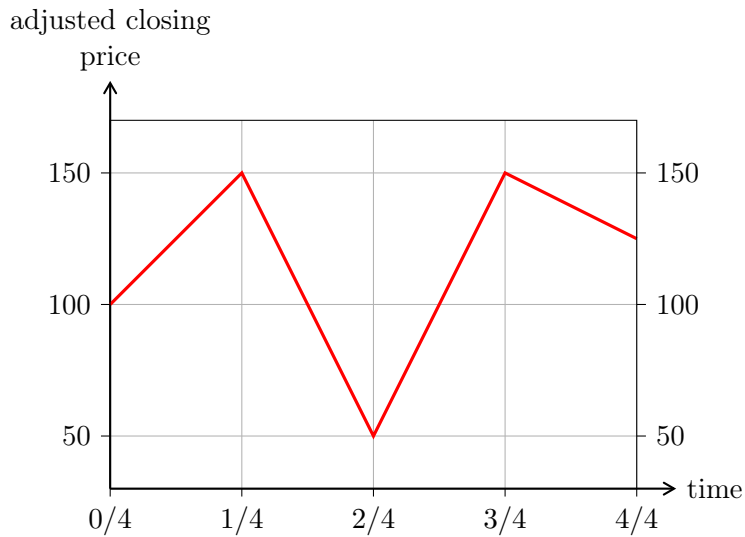
\begin{figure}[H]
  \captionsetup{margin = {-1.5cm,1.5cm}}
  \caption{\label{mean-return-problem}%
    The mean periodic return/discount problem
  }
\hspace*{1cm}%
\begin{tikzpicture}[scale=1.75, >={angle 60}, yscale=0.02]
  \foreach \x in {0,1,2,3,4}
    \draw[grid] (\x,30) -- (\x,170);
  \foreach \y in {50,100,150}
    \draw[grid] (0,\y) -- (4,\y);
  \draw[thick,->] (0,30) -- (4.3,30) node[right]{time};
  \foreach \x in {0,1,2,3,4}
    \draw (\x,30) -- (\x,27) node[below]{$\x/4$};
  \draw[thick,->] (0,30) -- (0,185) node[above]{
    \parbox{2.7cm}{\centering%
    adjusted closing price}};
  \foreach \y in {50,100,150}{
    \draw (0,\y) -- (-0.07,\y) node[left]{\y};
    \draw (4,\y) -- (4.07,\y) node[right]{\y};
  }
  \draw[red,very thick] (0,100) -- (1,150) -- (2,50) -- (3,150) -- (4,125);
  
  \draw (0,30) rectangle (4, 170);
\end{tikzpicture}

\begin{lstlisting}
%% hn2mv3.m - the mean periodic return/discount problem

%% quarterly adjusted closing prices
a     = [ 100, 150, 50, 150, 125 ];

%% annualized mean quarterly return
r     = a(2 : 5) ./ a(1 : 4) - 1;  % quarterly returns
e_r   = 4 * mean(r);        % e_r     =  1.6667
sig_r = 2 * std(r, 1);      % sig_r   =  2.0069

%% annualized mean quarterly discount
d     = 1 - a(1 : 4) ./ a(2 : 5);  % quarterly discounts
e_d   = 4 * mean(d);        % e_d     = -1.2000
sig_d = 2 * std(d, 1);      % sig_d   =  2.0580

%% annualized mean quarterly change in year-to-date return
df_0  = diff(a) / a(1);  % changes in year-to-date return
e_0   = 4 * mean(df_0);     % e_0     =  0.2500
sig_0 = 2 * std(df_0, 1);   % sig_0   =  1.5155

%% annualized mean quarterly change in date-to-end-of-year discount
df_1  = diff(a) / a(5);  % changes in date_to_end_of_year discount
e_1   = 4 * mean(df_1);     % e_1     =  0.2000
sig_1 = 2 * std(df_1, 1);   % sig_1   =  1.2124

%% return-discount relationaship as per "The Theory of Interest"
(1 + e_0) * (1 - e_1)       %         =  1

%% meaningless return-discount relationship
(1 + e_r) * (1 - e_d)       %         =  5.8667
\end{lstlisting}
\end{figure}

Figure \ref{mean-return-problem} on the next page and the
\verb!hn2domv3.m! script below it illustrate a serious problem with
mean periodic returns. This is a simple, artifical example, where a fund
gains 50\% in the first quarter of the year, loses 67\% in the second
quarter, gains 200\% in the third quarter, and loses 17\% in the forth
quarter.

An investor in the fund realizes that his fund has returned 25\% over
the year, but the trip has been terribly rocky; he decides to bail out.

Not so fast, his investment adviser tells him. Just add up the quarterly
returns:\\[0.5ex]
\hspace*{2cm}  $ 50\% - 67\% + 200\% - 17\% = 166\%. $\\[0.5ex]
You have avergaed averaged over 40\% per quarter. The fund may seem a
bit risky, but, in view of its history, you should \emph{expect} to
average 40\% per quarter next year as well. It's clear from the numbers.

Figure \ref{mean-return-problem} tells the whole story. Mean periodic
returns tend to accentuate the positive. Mean periodic discounts do just
do just the opposite.

\begin{defn}[Effective return and discount]\label{effective-return-discount}
  Let $a_0$ and $a_1$ be the the adjusted closing prices of a security
  on two different market days with $a_0$ occuring before $a_1$. Then
  the \emph{effective return} of the security over that period of time
  is defined as\\
    \hspace*{5cm}$r = (a_1 - a_0) / a_0,$\\
  and the \emph{effective discount} as\\
    \hspace*{5cm}$d = (a_1 - a_0) / a_1.$\\
  The equation\\
    \hspace*{5cm}$(1 + r)(1 - d) = 1$\\
  always holds, and $r$ and $d$ always
  have the same sign, positive, negative, or zero.
\end{defn}

Definition \ref{effective-return-discount} is a paraphrase of
definitions in \textbf{The Theory of Interest}, \cite{Kellison:2009aa}.

\vspace*{2ex}%
The means of periodic changes in year-to-date return and periodic
changes in date-to-end-of-year discount are the appropriate measures of
effective performance of a security over a year. In the the example of
Figure \ref{mean-return-problem}, the annualized mean of the changes
in year-to date return is\\[1ex]
  \hspace*{2cm}  $ e^0 = 50\% - 00\% + 100\% - 25\% = 25\%, $\\
and the annualized mean of changes in date-to-end-of-year discount is\\
  \hspace*{2cm}  $ e^1 = 40\% - 80\% + 80\% - 20\% = 20\%.$\\[1ex]
These annualized means do satisfy Definition \ref{effective-return-discount},
~$(1 + e^0)(1 - e^1) = 1$.

\begin{note*}
The $e^0$ and $e^1$ above correspond to the \lstinline!e_0! and
\lstinline!e_1! in the MATLAB code underneath Figure
\ref{mean-return-problem}. Likewise $e^r$ and $e^d$ correspond to
\lstinline!e_r! and \lstinline!e_d!.
\end{note*}

On the other hand, the MATLAB code shows that the mean return, $e^r$,
and the mean discount, $e^d$, have opposite signs, and
\mbox{$(1 + e^r)(1 - e^d) = 5.8667$}. These computations show that mean
returns and mean discounts are essentially incompatible with the theory
of interest. The mean return problem has been noted, for example, in
\cite[pp. 104-105]{Swensen:2009aa}.

\newpage

\newpage

SEC Rule 156 below might apply to the situation illustrated by 
Figure \ref{mean-return-problem} and the
\verb!hn2mv3.m! script which follows it.

\begin{verbatim}
     Rule 156: Investment Company Sales Literature 
     
     Under the federal securities laws, including section 17(a) of the
     Securities Act of 1933 (15 U.S.C. 77q(a)) and section 10(b) of the
     Securities Exchange Act of 1934 (15 U.S.C. 78j(b)) and Rule 10b-5
     thereunder (17 CFR Part 240), it is unlawful for any person,
     directly or indirectly, by the use of any means or instrumentality
     of interstate commerce or of the mails, to use sales literature
     which is materially misleading in connection with the offer or sale
     of securities issued by an investment company. Under these
     provisions, sales literature is materially misleading if it:
     
       1. Contains an untrue statement of a material fact or
       2. omits to state a material fact necessary in order to make a
          statement made, in the light of the circumstances of its use,
          not misleading. 
\end{verbatim}
\vspace*{-2ex}\hspace*{7cm}%
\href{https://www.investopedia.com/exam-guide/finra-series-6/marketing-presentations/securities-act-1933-rule-156.asp}
{Securities Act of 1933: Rule 156}\\

Rule 156 raises an interesting question. Is the use of mean-variance
analysis, as it appears to be practiced today, ``materially misleading''
when an investment company tells a client to ``expect'' a 160\% return
over the next year based on a 25\% total return over the past year? This
sort of reasoning reminds us of Mark Twain's analysis of the expected
shortening of the lower Mississippi due to the rounding of its bends
over time
(Appendix \ref{twain}).\\[1.0ex]

More realistically, consider the example of \ticker{FBT}. In Table
\ref{hn2mv1_table} we have seen that the annualized mean daily return of
\ticker{FBT} over 2014 was~
$e_{\sstick{FBT}}=e^r_{\sstick{FBT}}=42.45\%$. An investor with a
marginal knowledge of the theory of interest might ask his advisor what
the corresponding annualized mean daily discount was. If the advisor
were perplexed by this question, the investor could explain that to get
the annualized mean discount you simply replace the daily return
equation \eqref{daily_returns} by the daily discount equation
\begin{equation}\label{daily_discounts}
  d_i = 1 - a_{i-1}/a_i \quad(i = 1,\ldots,252)\hspace*{3.6cm}
\end{equation}
and sum the results. The advisor might then be mildly concerned by the
annualized average discount,~ $e^d_{\sstick{FBT}}=35.34\%$, if he were
told that returns and discounts over the same period of time are
supposed to satisfy the equation\\[0.5ex] \hspace*{3cm}$(1 + r)(1 - d) =
1$,\\[0.5ex] according to the theory of interest, but, in fact, $(1 +
e^r_{\sstick{FBT}})(1 - e^d_{\sstick{FBT}}) = 0.9211$.

\newpage
\section{How to do it -- the linear model}
\label{how-to-linear}

Our MATLAB script \verb!hn2mv1L.m! is a linear variant of the
\verb!hn2mv1a.m! script
of Section \ref{how-not-to}. We again remove $\mathbf{a}_{\sstick{CRP}}$
from $A$ and append the unattended portfolio vectors
$\mathbf{a}_{\sstick{UIP2}}$ and $\mathbf{a}_{\sstick{UIP3}}$ to the
result. Then we add the unattended, long-short portfolio
\ticker{ZNS}, with normalized adjusted closing price vector
\begin{equation}\label{aZNS1}\vspace*{-1ex}\hspace*{-3ex}%
  \mathbf{a}_{\sstick{ZNS}} = 1.25515 \cdot \mathbf{a}_{\sstick{FBT}}
  - 0.25515 \cdot \mathbf{a}_{\sstick{XBI}}\\[0.5ex]
\end{equation}
to $A$, so that $A$ becomes the $253 \times 6$ matrix\\[0.5ex]
\hspace*{4cm}%
$A = \left[\,\mathbf{a}_{\sstick{FBT}},\, \mathbf{a}_{\sstick{XBI}},\,
  \mathbf{a}_{\sstick{UIP}},\, \mathbf{a}_{\sstick{UIP2}},\,
  \mathbf{a}_{\sstick{UIP3}},\, \mathbf{a}_{\sstick{ZNS}}\,\right]$\\
of rank 2. Finally, we put $\mathbf{a}_{\sstick{CRP}}$ back into $A$,\\[-2ex]
\begin{equation}\label{bigA}
A = \left[\,\mathbf{a}_{\sstick{FBT}},\, \mathbf{a}_{\sstick{XBI}},\,
  \mathbf{a}_{\sstick{UIP}},\, \mathbf{a}_{\sstick{UIP2}},\,
  \mathbf{a}_{\sstick{UIP3}},\, \mathbf{a}_{\sstick{ZNS}},\,
  \mathbf{a}_{\sstick{CRP}}\,\right],
\end{equation}
and, since $\mathbf{a}_{\sstick{CRP}}$ is independent of the other
six columns of $A$, the rank of $A$ increases to 3.

When the lines\\[-1ex]
\hspace*{3ex}%
\begin{minipage}{13cm}
\begin{lstlisting}
%% get asset moments from adjusted closing prices
ptf = Portfolio;
ptf = estimateAssetMoments(ptf, A, 'dataformat', 'prices');
[mn, cv] = getAssetMoments(ptf);
\end{lstlisting}
\end{minipage}\\ of \verb!hn2mv1a.m! are replaced with the lines\\[-1ex]
\hspace*{3ex}
\begin{minipage}{13cm}
\begin{lstlisting}
%% get daily changes in year to date return
R_0 = diff(A / 100);       % divide by 100 so that A(1, :) == 1
%% get (annualized) asset moments
E_0 = sum(R_0)                   % total return
Sig_0 = sqrt(252) * std(R_0, 1); % standard deviation of return
V_0 = 252 * cov(R_0, 1);         % covariance of return
\end{lstlisting}
\end{minipage}\\
in \verb!hn2mv1L.m!, we arrive at the mean-variance results

\begin{table}[H]
  \centering
  \caption{\label{hn2mv1L_table}%
    Annualized results from the linear model}
$\arrayrulecolor{black}
\begin{array}{|l|rrrrrr|>{\columncolor{CRi}}r|}
  \hhline{~-------} 
  \multicolumn{1}{c|}{\rule{0mm}{4.0mm}}
  & \smtick{FBT}~ & \smtick{XBI}~~ & \smtick{UIP}~~
  & \smtick{UIP2}~ & \smtick{UIP3}~ & \smtick{ZNS}~\,\,
  & \smtick{CRP}\hspace*{1ex}\\\hline\rule{0mm}{4.2mm}%
  E^0 & 0.4754 & 0.4497 & 0.4690 & 0.4626 & 0.4562 & 0.4820 & 0.4732 \\
  \lgsig^0 & 0.3205 & 0.4050 & 0.3323 & 0.3510 & 0.3756 & 0.3163 & 0.3334 \\
  \hline
  \multicolumn{8}{c}{\text{covariance~} V^0~~~\rule{0mm}{5mm}}\\
  \hline\rule{0mm}{4mm}
  \smtick{FBT} & 0.1027 & 0.1131 & 0.1053 & 0.1079 & 0.1105 & 0.1001 & 0.1056 \\
  \smtick{XBI} & 0.1131 & 0.1641 & 0.1258 & 0.1386 & 0.1513 & 0.1001 & 0.1264 \\
  \smtick{UIP} & 0.1053 & 0.1258 & 0.1104 & 0.1156 & 0.1207 & 0.1001 & 0.1108 \\
  \smtick{UIP2} & 0.1079 & 0.1386 & 0.1156 & 0.1232 & 0.1309 & 0.1001 & 0.1160 \\
  \smtick{UIP3} & 0.1105 & 0.1513 & 0.1207 & 0.1309 & 0.1411 & 0.1001 & 0.1212 \\
  \smtick{ZNS} & 0.1001 & 0.1001 & 0.1001 & 0.1001 & 0.1001 & 0.1001 & 0.1003 \\
  \rowcolor{CRi}\hline\rule{0mm}{4mm}
  \smtick{CRP} & 0.1056 & 0.1264 & 0.1108 & 0.1160 & 0.1212 & 0.1003 & 0.1112 \\
  \hline 
\end{array}$
\end{table}

The covariance matrix $V^0$ has rank three, but the upper-left
$6\times6$ block has rank only two, since the four unattended portfolios
\ticker{UIP} through \ticker{ZNS} are affine combinations of the two funds
\ticker{FBT} and \ticker{XBI}. Moreover, the corresponding portion of
the total return matrix $E^0$ mirrors these affine combinations (in
contrast to the confusing order of the five mean return values in the
$E$ of Table \ref{hn2mv1_table}).

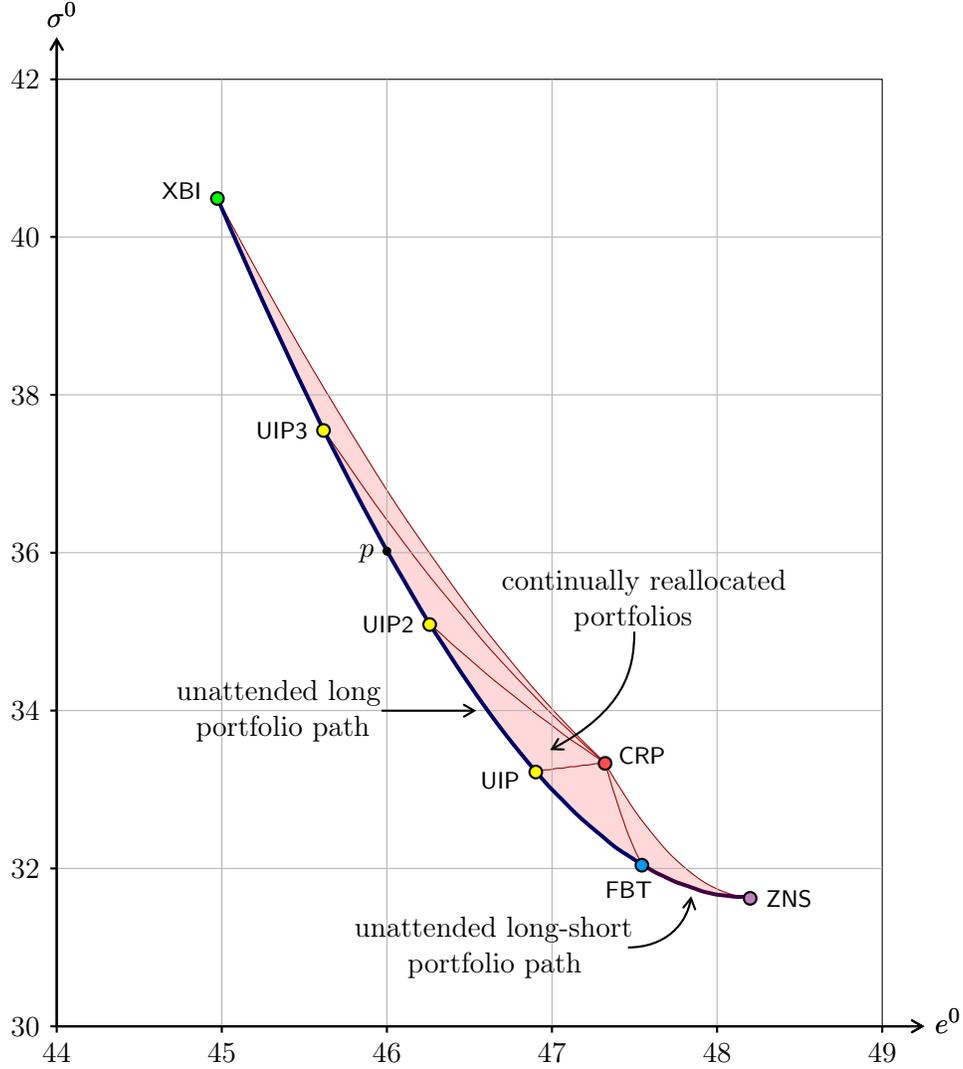
\begin{figure}[H]
  \centering
  \caption{\label{hn2mvL1_tikz}%
    Mean-variance analysis -- the linear model part 1
  }
\newcommand{\yscale}{3.75}
\begin{tikzpicture}[scale=28,yscale=\yscale,>={angle 60}]
  \setlength{\pointradius}{0.03mm}
  \setlength{\smallradius}{0.4\pointradius}

  \newcommand{\xL}{-0.32924}  
  \newcommand{\xH}{0.06271} 
  \newcommand{\dx}{0.07839}
  \newcommand{\yL}{0.30}
  \newcommand{\yH}{0.42}
  \newcommand{\dy}{0.02}

  \coordinate (FBT)  at (-0.0514,0.3205);
  \coordinate (XBI)  at (-0.2530,0.4050);
  \coordinate (UIP)  at (-0.1018,0.3323);
  \coordinate (UIP2) at (-0.1522,0.3510);
  \coordinate (UIP3) at (-0.2026,0.3756);
  \coordinate (ZNS)  at ( 0.0000,0.3163);
  \coordinate (CRP)  at (-0.0689,0.3334);
  \coordinate (P)    at (-0.1725,0.3603);
    
  \draw (\xL,\yL) rectangle (\xH,\yH);
  
  \draw[->,thick] (\xL,\yL) -- (\xH+0.02,\yL) node[above=2,right]{$e^0$};
  \draw[->,thick] (\xL,\yL) -- (\xL,\yH+0.02/\yscale) node[right=2,above]{$\sigma^0$};
  
  \draw[thick] (\xL,\yL) -- (\xL,\yL-0.003/\yscale) node[below]{44};
  \draw[thick] (\xL+\dx,\yL) -- (\xL+\dx,\yL-0.003/\yscale) node[below]{45};
  \draw[thick] (\xL+2*\dx,\yL) -- (\xL+2*\dx,\yL-0.003/\yscale) node[below]{46};
  \draw[thick] (\xL+3*\dx,\yL) -- (\xL+3*\dx,\yL-0.003/\yscale) node[below]{47};
  \draw[thick] (\xL+4*\dx,\yL) -- (\xL+4*\dx,\yL-0.003/\yscale) node[below]{48};
  \draw[thick] (\xH,\yL) -- (\xH,\yL-0.003/\yscale) node[below]{49};
  \draw[thick] (\xL,\yL) -- (\xL-0.003,\yL) node[left]{30};
  \draw[thick] (\xL,\yL+\dy) -- (\xL-0.003,\yL+\dy) node[left]{32};
  \draw[thick] (\xL,\yL+2*\dy) -- (\xL-0.003,\yL+2*\dy) node[left]{34};
  \draw[thick] (\xL,\yL+3*\dy) -- (\xL-0.003,\yL+3*\dy) node[left]{36};
  \draw[thick] (\xL,\yL+4*\dy) -- (\xL-0.003,\yL+4*\dy) node[left]{38};
  \draw[thick] (\xL,\yL+5*\dy) -- (\xL-0.003,\yL+5*\dy) node[left]{40};
  \draw[thick] (\xL,\yH) -- (\xL-0.003,\yH) node[left]{42};

  \filldraw[fill=CRi,draw=CR] plot[smooth] coordinates {
  (-0.2530,0.4050) 
  (-0.2467,0.4011) (-0.2404,0.3973) (-0.2341,0.3935) (-0.2278,0.3898)
  (-0.2215,0.3862) (-0.2152,0.3826) (-0.2089,0.3791) (-0.2026,0.3756)
  (-0.1963,0.3723) (-0.1900,0.3690) (-0.1837,0.3658) (-0.1774,0.3627)
  (-0.1711,0.3596) (-0.1648,0.3567) (-0.1585,0.3538) (-0.1522,0.3510)
  (-0.1459,0.3484) (-0.1396,0.3458) (-0.1333,0.3433) (-0.1270,0.3409)
  (-0.1207,0.3386) (-0.1144,0.3364) (-0.1081,0.3343) (-0.1018,0.3323)
  (-0.0955,0.3304) (-0.0892,0.3287) (-0.0829,0.3270) (-0.0766,0.3255)
  (-0.0703,0.3241) (-0.0640,0.3227) (-0.0577,0.3216) (-0.0514,0.3205)
  } --  plot[smooth] coordinates {
  (-0.0514,0.3205) 
  (-0.0482,0.3200) (-0.0450,0.3195) (-0.0418,0.3191) (-0.0386,0.3187)
  (-0.0354,0.3183) (-0.0321,0.3180) (-0.0289,0.3177) (-0.0257,0.3174)
  (-0.0225,0.3171) (-0.0193,0.3169) (-0.0161,0.3167) (-0.0129,0.3166)
  (-0.0096,0.3165) (-0.0064,0.3164) (-0.0032,0.3164) ( 0.0000,0.3163)
  } --  plot[smooth] coordinates {
  ( 0.0000,0.3163)  
  (-0.0043,0.3164) (-0.0086,0.3167) (-0.0129,0.3171) (-0.0172,0.3176)
  (-0.0215,0.3182) (-0.0258,0.3190) (-0.0301,0.3199) (-0.0344,0.3209)
  (-0.0387,0.3220) (-0.0430,0.3233) (-0.0473,0.3247) (-0.0517,0.3262)
  (-0.0560,0.3278) (-0.0603,0.3296) (-0.0646,0.3314) (-0.0689,0.3334)
  } --  plot[smooth] coordinates {
  (-0.0689,0.3334) 
  (-0.0804,0.3364) (-0.0919,0.3396) (-0.1034,0.3430) (-0.1149,0.3467)
  (-0.1264,0.3505) (-0.1379,0.3546) (-0.1494,0.3589) (-0.1609,0.3633)
  (-0.1724,0.3679) (-0.1839,0.3728) (-0.1954,0.3778) (-0.2069,0.3829)
  (-0.2184,0.3882) (-0.2300,0.3937) (-0.2415,0.3993) (-0.2530,0.4050)
  };
  
  \draw[CR] plot[smooth] coordinates {
  (-0.0689,0.3334) 
  (-0.0678,0.3324) (-0.0667,0.3314) (-0.0656,0.3304) (-0.0645,0.3295)
  (-0.0634,0.3286) (-0.0623,0.3277) (-0.0612,0.3268) (-0.0601,0.3260)
  (-0.0591,0.3252) (-0.0580,0.3244) (-0.0569,0.3237) (-0.0558,0.3230)
  (-0.0547,0.3223) (-0.0536,0.3217) (-0.0525,0.3211) (-0.0514,0.3205)
  };
  \draw[CR] plot[smooth] coordinates {
  (-0.0689,0.3334) 
  (-0.0709,0.3333) (-0.0730,0.3333) (-0.0750,0.3332) (-0.0771,0.3331)
  (-0.0792,0.3331) (-0.0812,0.3330) (-0.0833,0.3329) (-0.0853,0.3329)
  (-0.0874,0.3328) (-0.0895,0.3327) (-0.0915,0.3327) (-0.0936,0.3326)
  (-0.0956,0.3325) (-0.0977,0.3325) (-0.0998,0.3324) (-0.1018,0.3323)
  };
  \draw[CR] plot[smooth] coordinates {
  (-0.0689,0.3334) 
  (-0.0741,0.3343) (-0.0793,0.3353) (-0.0845,0.3362) (-0.0897,0.3372)
  (-0.0949,0.3383) (-0.1001,0.3393) (-0.1053,0.3404) (-0.1105,0.3415)
  (-0.1157,0.3426) (-0.1209,0.3437) (-0.1262,0.3449) (-0.1314,0.3461)
  (-0.1366,0.3473) (-0.1418,0.3485) (-0.1470,0.3498) (-0.1522,0.3510)
  };
  \draw[CR] plot[smooth] coordinates {
  (-0.0689,0.3334) 
  (-0.0772,0.3353) (-0.0856,0.3374) (-0.0939,0.3395) (-0.1023,0.3418)
  (-0.1107,0.3441) (-0.1190,0.3465) (-0.1274,0.3490) (-0.1357,0.3517)
  (-0.1441,0.3544) (-0.1524,0.3571) (-0.1608,0.3600) (-0.1692,0.3630)
  (-0.1775,0.3660) (-0.1859,0.3692) (-0.1942,0.3724) (-0.2026,0.3756)
  };
  
  \draw[grid] (\xL+\dx,\yL) -- (\xL+\dx,\yH);
  \draw[grid] (\xL+2*\dx,\yL) -- (\xL+2*\dx,\yH);
  \draw[grid] (\xL+3*\dx,\yL) -- (\xL+3*\dx,\yH);
  \draw[grid] (\xL+4*\dx,\yL) -- (\xL+4*\dx,\yH);
  \draw[grid] (\xL,\yL+\dy) -- (\xH,\yL+\dy);
  \draw[grid] (\xL,\yL+2*\dy) -- (\xH,\yL+2*\dy);
  \draw[grid] (\xL,\yL+3*\dy) -- (\xH,\yL+3*\dy);
  \draw[grid] (\xL,\yL+4*\dy) -- (\xH,\yL+4*\dy);
  \draw[grid] (\xL,\yL+5*\dy) -- (\xH,\yL+5*\dy);
  
  \draw[XF,line width=1.5pt] plot[smooth] coordinates {
  (-0.2530,0.4050) 
  (-0.2467,0.4011) (-0.2404,0.3973) (-0.2341,0.3935) (-0.2278,0.3898)
  (-0.2215,0.3862) (-0.2152,0.3826) (-0.2089,0.3791) (-0.2026,0.3756)
  (-0.1963,0.3723) (-0.1900,0.3690) (-0.1837,0.3658) (-0.1774,0.3627)
  (-0.1711,0.3596) (-0.1648,0.3567) (-0.1585,0.3538) (-0.1522,0.3510)
  (-0.1459,0.3484) (-0.1396,0.3458) (-0.1333,0.3433) (-0.1270,0.3409)
  (-0.1207,0.3386) (-0.1144,0.3364) (-0.1081,0.3343) (-0.1018,0.3323)
  (-0.0955,0.3304) (-0.0892,0.3287) (-0.0829,0.3270) (-0.0766,0.3255)
  (-0.0703,0.3241) (-0.0640,0.3227) (-0.0577,0.3216) (-0.0514,0.3205)
  };
  \draw[FZ,line width=1.5pt] plot[smooth] coordinates {
  (-0.0514,0.3205) 
  (-0.0482,0.3200) (-0.0450,0.3195) (-0.0418,0.3191) (-0.0386,0.3187)
  (-0.0354,0.3183) (-0.0321,0.3180) (-0.0289,0.3177) (-0.0257,0.3174)
  (-0.0225,0.3171) (-0.0193,0.3169) (-0.0161,0.3167) (-0.0129,0.3166)
  (-0.0096,0.3165) (-0.0064,0.3164) (-0.0032,0.3164) ( 0.0000,0.3163)
  };
  
  \draw[->,thick] (\xL,\yL) -- (\xH+0.02,\yL) node[above=2,right]{$e^0$};
  \draw[->,thick] (\xL,\yL) -- (\xL,\yH+0.02/\yscale) node[right=2,above]{$\sigma^0$};

  \node at (-0.05,0.354) {\parbox{3.8cm}{continually reallocated\\
    \hspace*{5ex} portfolios}};
  \node at (-0.21,0.34)  {\parbox{3.5cm}{unattended long\\
    \hspace*{0.8ex} portfolio path}};
  \node at (-0.12,0.31)  {\parbox{3.8cm}{unattended long-short\\ 
    \hspace*{3.5ex} portfolio path}};
  
  \draw[line width=0.8pt,->] (-0.055,0.35) .. controls (-0.055,0.345)
    and (-0.06,0.340) .. (-0.095,0.335);
  \draw[line width=0.8pt,->] (-0.175,0.34) -- (-0.130,0.34);
  \draw[line width=0.8pt,->] (-0.058,0.31) ..  controls (-0.04,0.31)
    and (-0.03,0.313) .. (-0.028,0.3165);

  \filldraw[fill=FBTv,thick] (FBT) circle[x radius=\pointradius,
    y radius=\pointradius/\yscale]node[left=5,below=2]{\smtick{FBT}};
  \filldraw[fill=XBIv,thick] (XBI) circle[x radius=\pointradius,
    y radius=\pointradius/\yscale]node[above=3,left=2]{\smtick{XBI}};
  \filldraw[fill=UIPv,thick] (UIP) circle[x radius=\pointradius,
    y radius=\pointradius/\yscale]node[below=3,left=2]{\smtick{UIP}};;
  \filldraw[fill=UIPv,thick] (UIP2) circle[x radius=\pointradius,
    y radius=\pointradius/\yscale]node[left=2]{\smtick{UIP2}};
  \filldraw[fill=UIPv,thick] (UIP3) circle[x radius=\pointradius,
    y radius=\pointradius/\yscale]node[left=2]{\smtick{UIP3}};
  \filldraw[fill=ZNSv,thick] (ZNS) circle[x radius=\pointradius,
    y radius=\pointradius/\yscale]node[right=2]{\smtick{ZNS}};;
  \filldraw[fill=CRPv,thick] (CRP) circle[x radius=\pointradius,
    y radius=\pointradius/\yscale]node[above=3,right=1]{\smtick{CRP}};
  \fill[black] (P) circle[x radius=0.7\pointradius,
    y radius=0.7\pointradius/\yscale]node[below=1,left=1]{$p$};
\end{tikzpicture}
\end{figure}

\newpage
Figure \ref{hn2mvL1_tikz} shows the obtainable $(e^0,\sigma^0)$
corresponding to Table \ref{hn2mv1L_table}. This nonlinear triangle is
the $(e^0,\sigma^0)$-image of the triangle in $\mathds{R}^{252}$ with vertices
$\mathbf{r}^0_{\sstick{XBI}}, \mathbf{r}^0_{\sstick{ZNS}}$, and 
$\mathbf{r}^0_{\sstick{CRP}}$.

\vspace*{0.5cm}%
\begin{note*}
We mentioned the theory of interest in the last section and the
relationship between discount and return. To continue this discusion let
$a_i (i = 0, 1,\dots,n)$ be adjusted closing prices of a given security
over $n$ successive investment periods. Then the total return, $e^0$,
and the total discount, $e^1$, of the security over this time interval
are given by
\[ e^0 = \sum_{i=1}^n (a_i - a_{i-1})/a_0 = (a_n - a_0)/a_0
\quad\text{and}\quad e^1 = \sum_{i=1}^n (a_i - a_{i-1})/a_n
= (a_n - a_0)/a_n \,.\]
\end{note*}
Thus $e^0$ and $e^1$ conform to the return-discount requirement of the
theory of interest,\[
  (1 + e^0)(1 - e^1) = \frac{a_n}{a_0} \frac{a_0}{a_n} = 1\,,
\]
but a corresponding summand pair only conforms by accident.
\newpage
\subsection{The linear model -- part 2}\label{linear2}

The covariance matrix $V^0$ of Table \ref{hn2mv1L_table} is the
\href{https://en.wikipedia.org/wiki/Gramian_matrix}{Gram matrix},
$V^0 = (Z^0)^T Z^0$, of the $252\times7$ risk vector matrix
\begin{equation}\label{Z^0}
  Z^0 = R^0 - \bm{1}_{252}\,(E^0 / 252)\,,\\[-2ex]
\end{equation}
or\\\hspace*{4.4cm}\lstinline!Z_0 = R_0 - ones(252, 1) * (E_0 / 252)!\\[1ex]
in MATLAB code.
The columns of $Z^0$ represent pure risk in that the sum of each
column is zero (= zero total return).

Table \ref{hn2mv1L_table} and
Figure \ref{hn2mvL1_tikz} are summaries of the 2014 adjusted closing
price histories of \ticker{FBT}, \ticker{XBI}, \ticker{UIP}, \ticker{UIP2},
\ticker{UIP3}, \ticker{ZNS}, and \ticker{CRP}. $E^0$ and $Z^0$ \emph{are}
the complete histories split into their return and risk parts.

For example, let $e^0$ and $\mathbf{z^0}$ be the total return and
risk vector of any one of the seven funds in Table
\ref{hn2mv1L_table}. Then Algorithm \ref{a_from_ez} will compute
the the normalized 2014 adjusted closing price history of
this fund (starting from \$100 at the close of 2013-12-31).\\
\hspace*{3cm}%
\begin{minipage}{7cm}
\begin{algorithm}[H]
  \caption{\label{a_from_ez}%
    To compute the adjusted closing price vector $\mathbf{a}$
    from $e^0$ and $\mathbf{z}^0$
  }
\begin{tikzpicture}[scale=1]
\node[right] at (0,0){ \hspace*{1cm}\parbox{4cm}{\vspace*{-0.4cm}
$\mathbf{r}^0 = \mathbf{z}^0 + (e^0/252)\, \bm{1}_{252};\\
a^0_0 = 1;\\
\text{for}~ i = 1,\ldots, 252\\
\hspace*{3.5ex} a^0_i = a^0_{i - 1} + r^0_i\,;\\[0.5ex]
\hspace*{0.1ex}\mathbf{a} = 100 * \mathbf{a}^0;
$}};
\draw (1.2, 0.05) -- (1.2, -0.4) -- (1.6, -0.4);\\
\end{tikzpicture}
\end{algorithm}
\end{minipage}

Our MATLAB code that illustrates this linear model section is
organized into four scripts.

\hspace*{3ex}%
\begin{minipage}{14.5cm}
\verb!hn2mv1L.m! ~\,-- compute the seven fund mean-variance
    table $E^0, V^0$ for the linear model\\
\verb!hn2mv1L1.m! -- construct (and save) the orthogonal
    $U, E^0, Z^0$  system (\verb!UEZ2014.mat!)\\
\verb!hn2mv1L2.m! -- generate an adjusted closing price
    history $A$ from the orthogonal system\\
\verb!hn2mv1L3.m! -- construct the seven fund $E^0, Z^0$
    table corresponding to the orthogonal\\\hspace*{2.3cm}
    system
\end{minipage}

We have already described the first script, \verb!hn2mv1L.m!.
The second script, \verb!hn2mv1L1.m!, takes the risk matrix
$Z^0$ \eqref{Z^0} apart orthogonally,
\begin{equation}\label{UEZ}
  Z^0 = U \widetilde{Z}^0,\quad
  \widetilde{Z}^0 = U^T Z^0,\quad
  U^T U = I,\quad
  U=\left[\mathbf{u}_x,\mathbf{u}_y,\mathbf{u}_z\right]
  \in\mathds{R}^{252\times3},
\end{equation}
\vspace*{-2ex}where $U$ is defined by
\begin{alignat}{4}
  \mathbf{v}_x &= \mathbf{z}^0_{\sstick{FBT}} \notag
    - \mathbf{z}^0_{\sstick{XBI}},\quad
  &\mathbf{u}_x &= \mathbf{v}_x / \| \mathbf{v}_x \|, \\
  \label{vuxyz}
  \mathbf{v}_y &= \mathbf{z}^0_{\sstick{FBT}}
    - \mathbf{u}_x (\mathbf{u}_x^T \mathbf{z}^0_{\sstick{FBT}}),\quad
  &\mathbf{u}_y &= \mathbf{v}_y / \| \mathbf{v}_y \|, \\
  \mathbf{v}_z &= \mathbf{z}^0_{\sstick{CRP}} \notag
    - \left[\mathbf{u}_x, \mathbf{u}_y\right]
    (\left[\mathbf{u}_x, \mathbf{u}_y\right]^T 
    \mathbf{z}^0_{\sstick{CRP}}),\quad
  &\mathbf{u}_z &= \mathbf{v}_z / \| \mathbf{v}_z \|,
\end{alignat}\\[-4ex]
and the resulting $\widetilde{Z}^0$ is
\begin{equation}\label{Z0_7}
\begin{array}{l|rrrrrr|>{\columncolor{CRi}}r|l}
  \multicolumn{1}{c}{\rule{0mm}{4.0mm}}
  & \smtick{FBT}~ & \smtick{XBI}~~ & \smtick{UIP}~~
  & \smtick{UIP2}~ & \smtick{UIP3}~ 
  & \multicolumn{1}{r}{\smtick{ZNS}~\,\,}
  & \multicolumn{1}{r}{\smtick{CRP}\hspace*{1ex}}\\
  \hhline{~-------~}\rule{0mm}{4.2mm}%
  & -0.0514 & -0.2530 & -0.1018 & -0.1522 & -0.2026 & 0~~ & -0.1030 & x\\
  \widetilde{Z}^0 = & 0.3163 & 0.3163 & 0.3163 & 0.3163 & 0.3163 & 0.3163 & 0.3171 & y \\
  & 0~~ & 0~~ & 0~~ & 0~~ & 0~~ & 0~~ & 0.0024 & z \\
  \cline{2-8}
\end{array}.
\end{equation}

\newpage
The orthonormal matrix $U$ and the \ticker{FBT}, \ticker{XBI},
\ticker{ZSN}, \ticker{CRP} columns (1, 2, 6, 7) of $E^0$ (Table \ref{hn2mv1L_table})
and $\widetilde{Z}^0$ \eqref{Z0_7} are saved as \lstinline{U}, \lstinline{E_0}, and
\lstinline{Z_0} in the MATLAB file \verb!UEZ2014.mat!. MATLAB Example
\ref{matlab3} verifies the contents this file.\\[-2ex]

\begin{matlab}[H]
  \centering
  \caption{\label{matlab3}%
    Linear mv-analysis -- check UEZ2014.mat
  }
\vspace*{-3ex}%
\begin{lstlisting}
%% hn2mv1L1check.m  - Linear mv-analysis
%                     Check UEZ2014.mat

load UEZ2014.mat;  % U  E_0  Z_0  dates  funds  legend
%{
legend =
  10 x 60 char array
  'U:  252 x 3 orthonormal matrix of risk vectors u: sum(u) = 0'
  'E_0:  1 x 4 matrix of total 2014 returns                    '
  'Z_0:  3 x 4 matrix of risk vector coordinates               '
  'dates: 252 x 10 string array of 2014 market days            '
  'funds: 4 x 3 string array of fund symbols                   '
  '         FBT - First Trust Biotechnology Index Fund         '
  '         XBI - SPDR S&P Biotech ETF                         '
  '         ZNS - Unattended long-short FBT-XBI portfolio      '
  '         CRP - Continually reallocated FBT-XBI portfolio    '
  'legend:  the above description                              '
%}

%% risk and orthogonality of U
%{
mean(U) = 1.0e-15 * [0.0039 -0.0007  0.4282]  % all risk
norm(U' * U - eye(3)) =  1.7697e-14           % orthonormal
%}

%% display E_0, Z_0, Sig_0, and V_0
%{
      FBT       XBI       ZNS       CRP
E_0 =
    0.4754    0.4497    0.4820    0.4732
Z_0 =
   -0.0514   -0.2530         0   -0.1030
    0.3163    0.3163    0.3163    0.3171
         0         0         0    0.0024
Sig_0 = sqrt(sum(Z_0 .^ 2)) =
    0.3205    0.4050    0.3163    0.3334
V_0 = Z_0' * Z_0 =
    0.1027    0.1131    0.1001    0.1056
    0.1131    0.1641    0.1001    0.1264
    0.1001    0.1001    0.1001    0.1003
    0.1056    0.1264    0.1003    0.1112
%}
\end{lstlisting}
\end{matlab}

\newpage

Table \ref{hn2mv1L_table2} below contains \emph{all} of the information
in Table \ref{hn2mv1L_table} in a more compact, geometric form. The
computation ~$V^0 = (Z^0)^T Z^0$~ reproduces the $V^0$ of
Table \ref{hn2mv1L_table}

\begin{table}[H]
  \centering
  \caption{\label{hn2mv1L_table2}%
    Annualized results from the linear model -- part 2}
  \vspace*{-1ex}$
\begin{array}{|l|rrrrrr|>{\columncolor{CRi}}r|l}
  \hhline{~-------~}\multicolumn{1}{c|}{\rule{0mm}{4.0mm}}
  & \smtick{FBT}~ & \smtick{XBI}~~ & \smtick{UIP}~~
  & \smtick{UIP2}~ & \smtick{UIP3}~ & \smtick{ZNS}~\,\,
  & \smtick{CRP}\hspace*{1ex}\\
  \hhline{--------~}\rule{0mm}{4.2mm}%
  E^0 & 0.4754 & 0.4497 & 0.4690 & 0.4626 & 0.4562 & 0.4820 & 0.4732 \\
  \lgsig^0 & 0.3205 & 0.4050 & 0.3323 & 0.3510 & 0.3756 & 0.3163 & 0.3334 \\
  \hhline{--------~}\rule{0mm}{4mm}
  & -0.0514 & -0.2530 & -0.1018 & -0.1522 & -0.2026 & 0~~ & -0.1030 & x\\
  Z^0 & 0.3163 & 0.3163 & 0.3163 & 0.3163 & 0.3163 & 0.3163 & 0.3171 & y \\
  & 0~~ & 0~~ & 0~~ & 0~~ & 0~~ & 0~~ & 0.0024 & z \\
  \cline{1-8}
\end{array}$
\end{table}

\vspace*{1ex}
The $E^0$ and $\lgsig^0$ rows of Table \ref{hn2mv1L_table2} are exactly
the same as those of Table \ref{hn2mv1L_table}, but the $Z^0$ of Table
 \ref{hn2mv1L_table2} replaces the $V^0 = (Z^0)^T Z^0$ of Table
 \ref{hn2mv1L_table} and, in each column, $\sigma^0 = \|\mathbf{z}^0\|$.

\vspace*{2ex}
\begin{minipage}{7.7cm}
  Figure \ref{hn2mvLTikZ} shows the $xy$-plane in the risk hyperplane
  $\{\mathbf{z}\in\mathds{R}^{252}:
  \texttt{mean}(\mathbf{z})= 0\}$. It exactly reflects the $Z^0$ data in Table
  \ref{hn2mv1L_table2}. Of course $\mathbf{z}^0_{\sstick{CRP}}$ is not
  in $xy$-plane as evidenced by its nonzero $z$-coordinate and the fact
  that its projection onto the $xy$-plane is linearly incompatible with
  its total 2014 return, $e^0_{\sstick{CRP}}$.\\
  
  The illustrative unattended portfolio\\\hspace*{0.9cm}%
  $p = 0.39946\times\smtick{FBT} + 0.60054\times\smtick{XBI}$\\
  (at 2013-12-31 closing prices)
  in Figure \ref{hn2mvL1_tikz} and
  \ref{hn2mvLTikZ} had total 2014 return $e^0_p = 0.4600$ with
  $x_p = -0.1725$ and $\sigma^0_p = 0.3603$.
\end{minipage}\hfill%
\begin{minipage}{6.4cm}
\begin{figure}[H]
  \caption{\label{hn2mvLTikZ}%
    The $xy$-plane in risk space
  }
\begin{tikzpicture}[scale=16,>={angle 60}]
  \setlength{\pointradius}{0.06mm}
  \setlength{\smallradius}{0.4\pointradius}

  \newcommand{\xL}{-0.32924}  
  \newcommand{\xH}{0.06271} 
  \newcommand{\dx}{0.07839}
  \newcommand{\yscale}{3.75}
    
  \coordinate (FBT)  at (-0.0514,0.3163);
  \coordinate (XBI)  at (-0.2530,0.3163);
  \coordinate (UIP)  at (-0.1018,0.3163);
  \coordinate (UIP2) at (-0.1522,0.3163);
  \coordinate (UIP3) at (-0.2026,0.3163);
  \coordinate (ZNS)  at (0,0.3163);        
  \coordinate (CRP)  at (-0.1030,0.3171);  
  \coordinate (P)    at (-0.1725,0.3163);
  \coordinate (FBTe)  at (-0.0514,-0.08);
  \coordinate (XBIe)  at (-0.2530,-0.08);
  \coordinate (UIPe)  at (-0.1018,-0.08);
  \coordinate (UIP2e) at (-0.1522,-0.08);
  \coordinate (UIP3e) at (-0.2026,-0.08);
  \coordinate (ZNSe)  at (0,-0.08);        
  \coordinate (CRPe)  at (-0.069,-0.08);  
  \coordinate (Pe)    at (-0.1725,-0.08);  
  
  \draw[grid,step=0.05cm] (-0.301,0) grid (0,0.4);
  
  \draw[->,thick] (-0.31,0) -- (0.03,0) node[right]{$x$};
  \draw[->,thick] (0,0) -- (0,0.43) node[above]{$y$};
  \draw[->,thick] (-0.31,-0.08) -- (0.03,-0.08) node[above=2,right]{$e^0$};
  
  \foreach \x in {-0.3,-0.2,-0.1}
    \draw[thick] (\x,0) -- (\x,-0.01) node[below]{\x};
  \foreach \y in {0.1,0.2,0.3,0.4}
    \draw[thick] (0,\y) -- (0.01,\y) node[right]{\y};
  \draw[thick] (-0.251,-0.08) -- (-0.251,-0.09) node[below]{0.45};
  \draw[thick] (-0.173,-0.08) -- (-0.173,-0.09) node[below]{0.46};
  \draw[thick] (-0.094,-0.08) -- (-0.094,-0.09) node[below]{0.47};
  \draw[thick] (-0.016,-0.08) -- (-0.016,-0.09) node[below]{0.48};

  \draw[XF,very thick] (XBI) -- (FBT);
  \draw[FZ,very thick] (FBT) -- (ZNS);
  
  \draw (0,0) -- node[right=4,above]{$\sigma^0_p$} (P);
  
  \node at (-0.10,0.375){$\mathbf{z}^0_{\sstick{CRP}}$\,projection};
  \draw[->] (-0.11,0.363) -- (-0.104,0.322);
  
  \filldraw[fill=FBTv,thick] (FBT) circle[radius=\pointradius]
    node[above=2]{\smtick{FBT}};
  \filldraw[fill=XBIv,thick] (XBI) circle[radius=\pointradius]
    node[left=2,above=2]{\smtick{XBI}};
  \filldraw[fill=UIPv,thick] (UIP) circle[radius=\pointradius]
    node[below=2]{\smtick{UIP}};
  \filldraw[fill=UIPv,thick] (UIP2) circle[radius=0.8\pointradius]
    node[right=2,above=2]{\smtick{UIP2}};
  \filldraw[fill=UIPv,thick] (UIP3) circle[radius=0.8\pointradius]
    node[left=1,above=2]{\smtick{UIP3}};
  \filldraw[fill=ZNSv,thick] (ZNS) circle[radius=\pointradius]
    node[above=3,right=2]{\smtick{ZNS}};
  \fill[black] (0,0) circle[radius=\pointradius] node[below=4]{$\bm{0}$};
  \fill[black] (CRP) circle[radius=\smallradius];
  \fill[black] (P) circle[radius=0.8\pointradius] node[left=3,below=1]{$p$};
  \filldraw[fill=FBTv,thick] (FBTe) circle[radius=\pointradius]
    node[above=2]{\smtick{FBT}};
  \filldraw[fill=XBIv,thick] (XBIe) circle[radius=\pointradius]
    node[left=2,above=2]{\smtick{XBI}};
  \filldraw[fill=UIPv,thick] (UIP3e) circle[radius=0.8\pointradius]
    node[left=1,above=2]{\smtick{UIP3}};
  \filldraw[fill=UIPv,thick] (UIP2e) circle[radius=0.8\pointradius]
    node[right=1,above=2]{\smtick{UIP2}};
  \filldraw[fill=UIPv,thick] (UIPe) circle[radius=\pointradius]
    node[left=1,above=2]{\smtick{UIP}};
  \filldraw[fill=ZNSv,thick] (ZNSe) circle[radius=\pointradius]
    node[above=2]{\smtick{ZNS}};
  \filldraw[fill=CRPv,thick] (CRPe) circle[radius=0.8\pointradius];
  \draw[<-] (-0.067,-0.088) -- (-.057,-0.122);
  \node at (-0.040,-0.130){$e^0_{\sstick{CPR}}$};
  \fill[black] (Pe) circle[radius=0.8\pointradius];
\end{tikzpicture}
\end{figure}
\end{minipage}

\newpage
\section{History: adjusted closing prices revisted}
\label{revisit_adj}

\begin{figure}[H]
  \captionsetup{width=10cm}
  \caption{\label{FXUZplot}%
    2013-12-31-normalized adjusted closing prices of two biotechnology
    ETFs and two portfolios in these ETFs
  }
  \centering
  \vspace*{-2ex}
\begin{tikzpicture}[scale=1.1, >={angle 60},
  xscale=0.047,yscale=0.14]   
  
  \foreach \y in {100,110,120,130,140,150}
    \draw[grid] (0,\y) -- (252,\y);
  \foreach \y in {90,100,110,120,130,140,150,160}
    \draw (252,\y) -- (255,\y) node[right]{\y};
  \foreach \x in {61,124,188}
    \draw[grid] (\x,90) -- (\x,160);
  \foreach \x in {0,61,124,188,252}
    \draw (\x,90) -- (\x,89);
  \node[below=1] at (0,89) {2013-12-31};
  \node[below=1] at (61,89) {2014-03-31};
  \node[below=1] at (124,89) {2014-06-30};
  \node[below=1] at (188,89) {2014-09-30};
  \node[below=1] at (252,89) {2014-12-31};
  \draw[thick] (0,90) rectangle (252,160);
  
  \filldraw[fill=white] (4.7,142.5) rectangle (200,157);  
  
  \draw[line width=3pt,FBTa] (15,154.5) -- (20,154.5)
    node[black,right=2]{\ticker{FBT} =
    First Trust Biotechnology Index ETF};
  \draw[line width=3pt,XBIa] (15,151.5) -- (20,151.5)
    node[black,right=2]{\ticker{XBI} \hspace*{0.7ex}=
    SPDR S\&P Biotech ETF};
  \draw[line width=3pt,ZNSa] (15,148.5) -- (20,148.5)
    node[black,right=2]{\ticker{ZNS} \hspace*{0.2ex}=
    Unattended long-short \ticker{FBT}-\ticker{XBI} portfolio};
  \draw[line width=3pt,CRPa] (15,145.2) -- (20,145.2)
    node[black,right=2]{\ticker{CRP} \hspace*{0.1ex}=
    Continually reallocated \ticker{FBT}-\ticker{XBI} portfolio};
  
  \draw[ZNSa,line width=1pt] plot coordinates {  (0,100.00) 
  (1,100.42) (2,99.79) (3,99.67) (4,101.29) (5,102.49)
  (6,102.15) (7,103.25) (8,101.04) (9,106.51) (10,106.59)
  (11,107.78) (12,107.52) (13,109.98) (14,109.43) (15,109.76)
  (16,105.70) (17,104.03) (18,105.64) (19,105.23) (20,108.47)
  (21,107.11) (22,103.85) (23,104.38) (24,103.68) (25,103.39)
  (26,106.96) (27,108.86) (28,110.14) (29,110.65) (30,111.92)
  (31,111.27) (32,114.22) (33,112.52) (34,114.91) (35,116.32)
  (36,117.45) (37,125.25) (38,122.21) (39,121.43) (40,119.30)
  (41,120.26) (42,122.51) (43,122.70) (44,119.88) (45,120.25)
  (46,120.78) (47,119.40) (48,120.38) (49,117.49) (50,116.31)
  (51,117.07) (52,120.98) (53,119.26) (54,119.18) (55,114.45)
  (56,111.87) (57,112.01) (58,110.06) (59,112.37) (60,107.73)
  (61,111.16) (62,114.11) (63,114.79) (64,111.54) (65,108.31)
  (66,108.73) (67,107.80) (68,111.75) (69,105.62) (70,103.21)
  (71,103.30) (72,104.36) (73,107.02) (74,106.97) (75,107.98)
  (76,110.90) (77,109.48) (78,109.10) (79,105.54) (80,105.99)
  (81,108.93) (82,110.01) (83,110.62) (84,109.08) (85,112.07)
  (86,110.86) (87,109.78) (88,108.18) (89,110.36) (90,112.99)
  (91,111.79) (92,112.51) (93,110.41) (94,110.11) (95,112.37)
  (96,110.37) (97,111.14) (98,112.67) (99,111.82) (100,113.76)
  (101,113.30) (102,113.79) (103,113.25) (104,113.12) (105,114.23)
  (106,115.71) (107,116.67) (108,116.15) (109,115.66) (110,115.31)
  (111,115.41) (112,115.80) (113,116.60) (114,116.77) (115,116.01)
  (116,116.76) (117,116.63) (118,117.65) (119,116.59) (120,118.13)
  (121,118.16) (122,118.18) (123,118.51) (124,119.01) (125,121.82)
  (126,122.15) (127,122.61) (128,120.34) (129,118.53) (130,120.54)
  (131,120.04) (132,121.39) (133,121.95) (134,120.06) (135,118.84)
  (136,116.17) (137,119.05) (138,118.88) (139,119.96) (140,121.32)
  (141,119.35) (142,119.00) (143,118.54) (144,121.31) (145,121.17)
  (146,118.33) (147,118.89) (148,119.97) (149,119.95) (150,120.34)
  (151,119.54) (152,120.45) (153,121.25) (154,121.32) (155,123.77)
  (156,125.47) (157,126.06) (158,127.22) (159,128.04) (160,128.90)
  (161,127.92) (162,128.44) (163,134.08) (164,134.87) (165,134.65)
  (166,134.37) (167,136.56) (168,137.05) (169,136.86) (170,134.82)
  (171,135.04) (172,136.20) (173,134.76) (174,136.96) (175,135.34)
  (176,134.12) (177,132.15) (178,134.72) (179,134.85) (180,136.00)
  (181,135.77) (182,135.38) (183,134.93) (184,138.54) (185,135.85)
  (186,136.85) (187,136.95) (188,135.66) (189,133.76) (190,132.96)
  (191,135.14) (192,133.44) (193,131.02) (194,134.91) (195,131.95)
  (196,133.11) (197,129.87) (198,128.58) (199,128.34) (200,129.91)
  (201,131.72) (202,133.00) (203,136.69) (204,136.60) (205,139.69)
  (206,141.48) (207,143.38) (208,144.39) (209,143.66) (210,145.79)
  (211,146.54) (212,146.95) (213,145.40) (214,143.23) (215,145.34)
  (216,143.22) (217,144.47) (218,145.49) (219,145.34) (220,145.35)
  (221,142.00) (222,142.29) (223,144.93) (224,144.76) (225,144.67)
  (226,145.54) (227,147.77) (228,147.76) (229,149.16) (230,149.30)
  (231,148.35) (232,150.18) (233,150.29) (234,149.42) (235,150.08)
  (236,153.71) (237,153.42) (238,150.31) (239,150.33) (240,148.00)
  (241,144.25) (242,142.13) (243,146.81) (244,151.79) (245,153.29)
  (246,151.38) (247,145.00) (248,146.98) (249,150.65) (250,151.10)
  (251,149.14) (252,148.20)
  };
  
  \draw[CRPa,line width=1pt] plot coordinates {  (0,100.00) 
  (2,100.45) (3,99.94) (4,99.41) (5,101.34) (6,103.11)
  (7,106.04) (8,109.12) (9,106.88) (10,110.49) (11,110.80)
  (12,112.46) (13,112.65) (14,115.16) (15,114.67) (16,114.86)
  (17,110.77) (18,107.81) (19,110.13) (20,109.34) (21,112.36)
  (22,110.47) (23,106.41) (24,107.35) (25,106.07) (26,105.61)
  (27,110.20) (28,112.64) (29,114.08) (30,114.50) (31,115.81)
  (32,114.59) (33,117.60) (34,115.78) (35,118.50) (36,120.36)
  (37,121.69) (38,127.64) (39,125.65) (40,125.38) (41,121.99)
  (42,122.44) (43,125.06) (44,124.99) (45,121.69) (46,121.68)
  (47,122.19) (48,120.85) (49,121.70) (50,119.03) (51,118.52)
  (52,118.67) (53,122.74) (54,121.49) (55,120.83) (56,115.94)
  (57,112.49) (58,112.35) (59,109.74) (60,111.51) (61,107.13)
  (62,110.67) (63,113.36) (64,113.43) (65,109.86) (66,106.15)
  (67,106.33) (68,105.90) (69,109.85) (70,103.40) (71,100.34)
  (72,99.72) (73,100.41) (74,103.11) (75,102.81) (76,104.38)
  (77,107.89) (78,105.87) (79,105.35) (80,101.64) (81,101.65)
  (82,104.88) (83,105.67) (84,106.54) (85,104.87) (86,107.39)
  (87,105.58) (88,104.40) (89,102.04) (90,104.41) (91,107.33)
  (92,106.24) (93,106.70) (94,104.90) (95,104.44) (96,106.64)
  (97,104.61) (98,105.02) (99,106.89) (100,106.75) (101,109.34)
  (102,108.93) (103,109.48) (104,108.72) (105,108.11) (106,108.85)
  (107,110.55) (108,112.01) (109,112.17) (110,114.59) (111,115.55)
  (112,115.25) (113,115.47) (114,115.77) (115,116.73) (116,116.38)
  (117,117.37) (118,117.38) (119,118.21) (120,117.28) (121,118.07)
  (122,118.20) (123,118.19) (124,118.74) (125,119.20) (126,121.72)
  (127,121.94) (128,122.26) (129,119.30) (130,116.42) (131,118.03)
  (132,117.20) (133,118.45) (134,118.90) (135,115.98) (136,114.59)
  (137,111.40) (138,114.52) (139,114.61) (140,115.55) (141,119.45)
  (142,117.75) (143,116.89) (144,115.87) (145,119.00) (146,119.61)
  (147,116.40) (148,116.36) (149,117.51) (150,117.82) (151,118.18)
  (152,117.04) (153,118.63) (154,119.91) (155,119.49) (156,121.99)
  (157,123.48) (158,123.87) (159,125.08) (160,125.42) (161,125.47)
  (162,123.97) (163,124.74) (164,129.66) (165,131.49) (166,131.13)
  (167,130.22) (168,132.19) (169,132.04) (170,132.16) (171,130.23)
  (172,129.97) (173,131.34) (174,129.59) (175,132.53) (176,131.72)
  (177,130.31) (178,127.99) (179,129.77) (180,130.50) (181,131.22)
  (182,130.73) (183,129.54) (184,128.94) (185,132.72) (186,130.20)
  (187,131.50) (188,131.92) (189,129.93) (190,128.16) (191,128.31)
  (192,130.39) (193,128.31) (194,125.69) (195,129.08) (196,125.84)
  (197,125.33) (198,123.14) (199,122.92) (200,124.45) (201,126.78)
  (202,127.69) (203,129.36) (204,132.08) (205,131.59) (206,135.35)
  (207,137.02) (208,138.13) (209,140.12) (210,139.08) (211,141.53)
  (212,141.34) (213,141.63) (214,140.25) (215,137.54) (216,139.87)
  (217,138.21) (218,140.37) (219,141.02) (220,141.38) (221,140.64)
  (222,137.56) (223,137.88) (224,140.40) (225,139.84) (226,140.63)
  (227,141.62) (228,144.26) (229,144.10) (230,145.82) (231,145.68)
  (232,143.49) (233,146.05) (234,146.17) (235,145.22) (236,146.62)
  (237,149.11) (238,150.63) (239,147.51) (240,147.66) (241,146.18)
  (242,141.47) (243,139.89) (244,145.31) (245,150.13) (246,151.58)
  (247,149.99) (248,143.11) (249,145.38) (250,148.90) (251,149.38)
  (252,147.59) (253,147.32)
  };

  \draw[XBIa,line width=1pt] plot coordinates {  (0,100.00) 
  (1,100.49) (2,100.17) (3,99.02) (4,101.40) (5,104.03)
  (6,111.85) (7,117.98) (8,115.71) (9,116.38) (10,117.03)
  (11,119.42) (12,120.32) (13,122.90) (14,122.50) (15,122.47)
  (16,118.35) (17,113.38) (18,116.79) (19,115.40) (20,118.08)
  (21,115.37) (22,110.09) (23,111.64) (24,109.47) (25,108.76)
  (26,114.90) (27,118.16) (28,119.83) (29,120.11) (30,121.49)
  (31,119.40) (32,122.50) (33,120.48) (34,123.69) (35,126.26)
  (36,127.89) (37,130.95) (38,130.58) (39,131.07) (40,125.75)
  (41,125.43) (42,128.61) (43,128.15) (44,124.12) (45,123.54)
  (46,124.00) (47,122.75) (48,123.39) (49,121.06) (50,121.55)
  (51,120.78) (52,125.08) (53,124.55) (54,123.02) (55,117.87)
  (56,113.14) (57,112.58) (58,108.98) (59,109.97) (60,105.99)
  (61,109.66) (62,111.96) (63,111.14) (64,107.12) (65,102.72)
  (66,102.53) (67,102.84) (68,106.79) (69,99.89) (70,95.90)
  (71,94.26) (72,94.43) (73,97.16) (74,96.51) (75,98.88)
  (76,103.20) (77,100.34) (78,99.62) (79,95.72) (80,95.09)
  (81,98.71) (82,99.08) (83,100.33) (84,98.48) (85,100.32)
  (86,97.69) (87,96.36) (88,92.97) (89,95.58) (90,98.89)
  (91,97.94) (92,98.06) (93,96.67) (94,96.00) (95,98.09)
  (96,96.05) (97,95.96) (98,98.28) (99,99.11) (100,102.61)
  (101,102.28) (102,102.91) (103,101.83) (104,100.55) (105,100.77)
  (106,102.76) (107,104.92) (108,106.03) (109,112.57) (110,115.43)
  (111,114.54) (112,114.52) (113,114.08) (114,116.21) (115,116.45)
  (116,117.78) (117,118.00) (118,118.56) (119,117.83) (120,117.49)
  (121,117.78) (122,117.71) (123,118.59) (124,119.00) (125,121.08)
  (126,121.13) (127,121.24) (128,117.26) (129,112.84) (130,113.85)
  (131,112.56) (132,113.66) (133,113.94) (134,109.55) (135,107.93)
  (136,104.03) (137,107.48) (138,107.92) (139,108.67) (140,116.12)
  (141,114.82) (142,113.22) (143,111.39) (144,115.05) (145,116.71)
  (146,112.98) (147,112.09) (148,113.32) (149,114.12) (150,114.41)
  (151,112.79) (152,115.35) (153,117.34) (154,116.20) (155,118.77)
  (156,119.93) (157,120.05) (158,121.32) (159,120.95) (160,119.83)
  (161,117.59) (162,118.72) (163,122.57) (164,125.87) (165,125.32)
  (166,123.52) (167,125.15) (168,124.10) (169,124.64) (170,122.88)
  (171,121.96) (172,123.60) (173,121.42) (174,125.39) (175,125.75)
  (176,124.06) (177,121.25) (178,121.91) (179,123.48) (180,123.59)
  (181,122.74) (182,120.42) (183,119.63) (184,123.61) (185,121.34)
  (186,123.04) (187,123.91) (188,120.95) (189,119.37) (190,120.84)
  (191,122.78) (192,120.19) (193,117.31) (194,119.95) (195,116.34)
  (196,113.52) (197,112.79) (198,114.05) (199,118.04) (200,121.43)
  (201,121.04) (202,123.25) (203,124.58) (204,123.51) (205,128.19)
  (206,129.68) (207,129.67) (208,133.00) (209,131.54) (210,134.45)
  (211,132.91) (212,133.02) (213,131.89) (214,128.44) (215,131.06)
  (216,130.06) (217,133.50) (218,133.62) (219,134.69) (220,132.87)
  (221,130.22) (222,130.58) (223,132.90) (224,131.80) (225,133.82)
  (226,134.97) (227,138.20) (228,137.80) (229,139.98) (230,139.44)
  (231,135.48) (232,139.05) (233,139.18) (234,138.12) (235,140.58)
  (236,141.42) (237,145.50) (238,142.38) (239,142.72) (240,142.48)
  (241,136.40) (242,135.61) (243,142.08) (244,146.63) (245,147.99)
  (246,146.89) (247,139.31) (248,141.99) (249,145.26) (250,145.79)
  (251,144.26) (252,144.97)
  };
  
  \draw[FBTa,line width=1pt] plot coordinates {  (0,100.00) 
  (1,100.43) (2,99.87) (3,99.54) (4,101.31) (5,102.80)
  (6,104.12) (7,106.25) (8,104.02) (9,108.51) (10,108.72)
  (11,110.15) (12,110.12) (13,112.61) (14,112.08) (15,112.34)
  (16,108.27) (17,105.93) (18,107.91) (19,107.30) (20,110.42)
  (21,108.79) (22,105.12) (23,105.85) (24,104.86) (25,104.48)
  (26,108.57) (27,110.75) (28,112.11) (29,112.58) (30,113.86)
  (31,112.92) (32,115.90) (33,114.14) (34,116.70) (35,118.34)
  (36,119.57) (37,126.41) (38,123.91) (39,123.39) (40,120.61)
  (41,121.31) (42,123.75) (43,123.81) (44,120.74) (45,120.92)
  (46,121.44) (47,120.08) (48,120.99) (49,118.21) (50,117.38)
  (51,117.82) (52,121.81) (53,120.34) (54,119.96) (55,115.15)
  (56,112.13) (57,112.13) (58,109.84) (59,111.88) (60,107.37)
  (61,110.86) (62,113.67) (63,114.05) (64,110.64) (65,107.17)
  (66,107.47) (67,106.79) (68,110.74) (69,104.45) (70,101.72)
  (71,101.46) (72,102.34) (73,105.02) (74,104.84) (75,106.13)
  (76,109.34) (77,107.62) (78,107.17) (79,103.54) (80,103.77)
  (81,106.85) (82,107.79) (83,108.53) (84,106.92) (85,109.69)
  (86,108.18) (87,107.05) (88,105.09) (89,107.36) (90,110.12)
  (91,108.98) (92,109.57) (93,107.62) (94,107.24) (95,109.47)
  (96,107.46) (97,108.05) (98,109.74) (99,109.24) (100,111.49)
  (101,111.06) (102,111.58) (103,110.93) (104,110.57) (105,111.49)
  (106,113.08) (107,114.28) (108,114.09) (109,115.03) (110,115.34)
  (111,115.24) (112,115.54) (113,116.09) (114,116.65) (115,116.10)
  (116,116.97) (117,116.91) (118,117.84) (119,116.84) (120,118.00)
  (121,118.08) (122,118.08) (123,118.53) (124,119.01) (125,121.67)
  (126,121.94) (127,122.33) (128,119.72) (129,117.38) (130,119.18)
  (131,118.52) (132,119.82) (133,120.32) (134,117.92) (135,116.62)
  (136,113.70) (137,116.70) (138,116.65) (139,117.66) (140,120.27)
  (141,118.43) (142,117.82) (143,117.09) (144,120.03) (145,120.27)
  (146,117.25) (147,117.50) (148,118.62) (149,118.76) (150,119.14)
  (151,118.17) (152,119.41) (153,120.45) (154,120.28) (155,122.75)
  (156,124.34) (157,124.83) (158,126.02) (159,126.60) (160,127.06)
  (161,125.82) (162,126.47) (163,131.74) (164,133.04) (165,132.75)
  (166,132.16) (167,134.24) (168,134.42) (169,134.37) (170,132.39)
  (171,132.38) (172,133.64) (173,132.05) (174,134.60) (175,133.39)
  (176,132.08) (177,129.94) (178,132.12) (179,132.54) (180,133.48)
  (181,133.12) (182,132.34) (183,131.82) (184,135.50) (185,132.90)
  (186,134.04) (187,134.30) (188,132.67) (189,130.83) (190,130.50)
  (191,132.62) (192,130.75) (193,128.23) (194,131.87) (195,128.78)
  (196,129.13) (197,126.39) (198,125.63) (199,126.25) (200,128.19)
  (201,129.55) (202,131.02) (203,134.23) (204,133.94) (205,137.35)
  (206,139.09) (207,140.59) (208,142.08) (209,141.20) (210,143.48)
  (211,143.77) (212,144.12) (213,142.66) (214,140.23) (215,142.44)
  (216,140.55) (217,142.24) (218,143.08) (219,143.18) (220,142.82)
  (221,139.61) (222,139.91) (223,142.48) (224,142.12) (225,142.47)
  (226,143.39) (227,145.82) (228,145.74) (229,147.30) (230,147.30)
  (231,145.74) (232,147.92) (233,148.03) (234,147.12) (235,148.15)
  (236,151.21) (237,151.81) (238,148.70) (239,148.79) (240,146.88)
  (241,142.66) (242,140.81) (243,145.85) (244,150.74) (245,152.21)
  (246,150.46) (247,143.84) (248,145.97) (249,149.55) (250,150.02)
  (251,148.15) (252,147.54)
  };
\end{tikzpicture}
\end{figure}
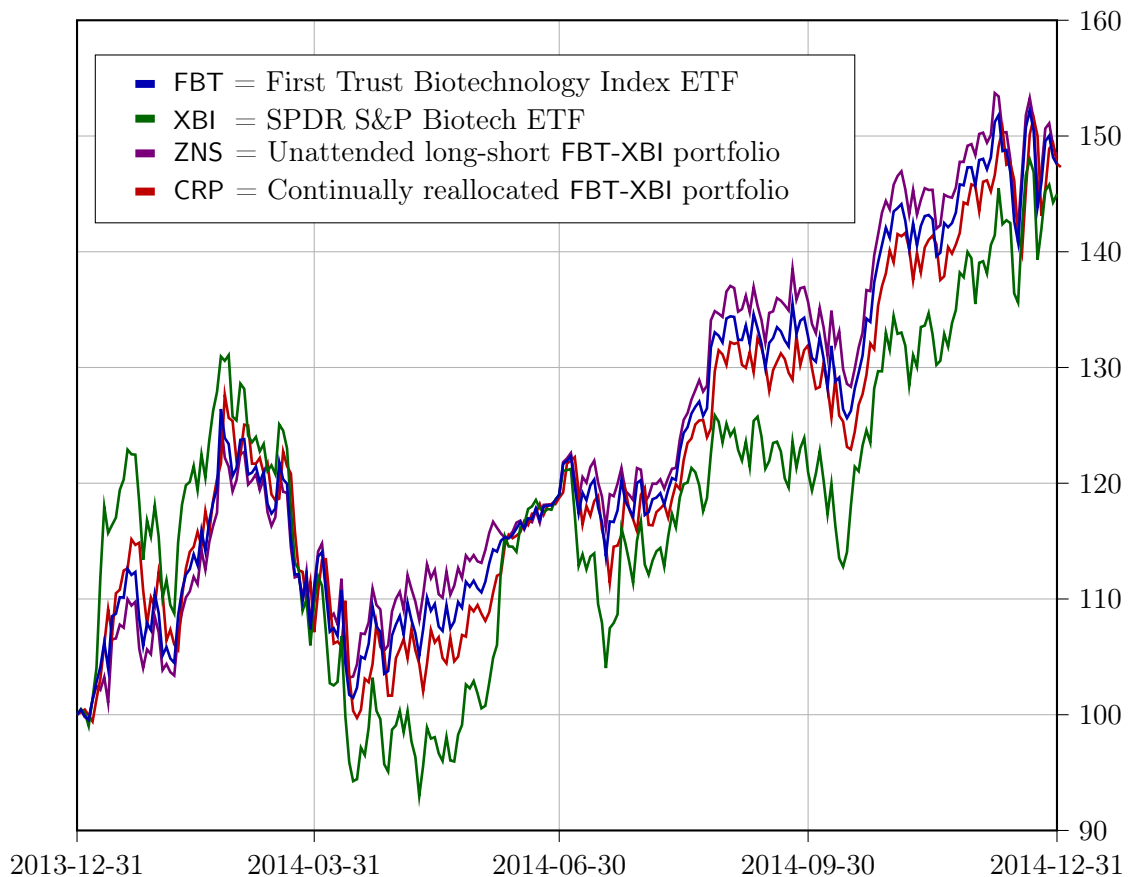

Let us close this article with a revised version of Figure
\ref{FXUCplot}. The revision, Figure \ref{FXUZplot}, shows the same
adjusted-closing-price history of the exchange traded funds \ticker{FBT}
and \ticker{XBI} and the continually reallocated portfolio \ticker{CRP},
but now the unattended long-short portfolio, \ticker{ZNS}, has replaced
\ticker{UIP}. Of the four funds and portfolios, \ticker{ZNS} (purple)
had the highest 2014 return with the least volatility.

These normalized adjusted closing prices were generated from
\verb!UEZ2014.mat! by the MATLAB script \verb!hn2mv1L3.m!. They are
recorded, to 5-decimal places (along with the prices of \ticker{UIP},
\ticker{UIP2}, and \ticker{UIP}3), in the comma-separated-value file
\verb!FXUZC7.csv!.

\newpage
Now the MATLAB script \verb!hn2mv1b.m!, with the $A$ of \ref{bigA} in
the
\\\hspace*{3ex}%
\lstinline!ptf = estimateAssetMoments(ptf, A, 'dataformat', 'prices');!\\
code produces\\[-3ex]
\begin{table}[H]
  \centering
  \caption{\label{hnmv1B_table}%
    Annualized results from the traditional nonlinear model}
$
\arrayrulecolor{black}
\begin{array}{|l|rrrrrr|>{\columncolor{CRi}}r|}
  \hhline{~-------} 
  \multicolumn{1}{c|}{\rule{0mm}{4.0mm}}
  & \smtick{FBT}~ & \smtick{XBI}~~ & \smtick{UIP}~~
  & \smtick{UIP2}~ & \smtick{UIP3}~ & \smtick{ZNS}~\,\,
  & \smtick{CRP}\hspace*{1ex}\\\hline\rule{0mm}{4.2mm}%
  E & 0.4245 & 0.4324 & 0.4235 & 0.4244 & 0.4274 & 0.4275 & 0.4265 \\
  \lgsig & 0.2656 & 0.3491 & 0.2777 & 0.2963 & 0.3203 & 0.2605 & 0.2783 \\
  \hline
  \multicolumn{8}{c}{\text{covariance~} V~~~\rule{0mm}{5mm}}\\
  \hline\rule{0mm}{4mm}
  \smtick{FBT} & 0.0705 & 0.0804 & 0.0729 & 0.0753 & 0.0778 & 0.0682 & 0.0730 \\
  \smtick{XBI} & 0.0804 & 0.1219 & 0.0905 & 0.1008 & 0.1112 & 0.0703 & 0.0908 \\
  \smtick{UIP} & 0.0729 & 0.0905 & 0.0771 & 0.0815 & 0.0859 & 0.0686 & 0.0773 \\
  \smtick{UIP2} & 0.0753 & 0.1008 & 0.0815 & 0.0878 & 0.0942 & 0.0691 & 0.0817 \\
  \smtick{UIP3} & 0.0778 & 0.1112 & 0.0859 & 0.0942 & 0.1026 & 0.0697 & 0.0862 \\
  \smtick{ZNS} & 0.0682 & 0.0703 & 0.0686 & 0.0691 & 0.0697 & 0.0678 & 0.0687 \\
  \rowcolor{CRi}\hline\rule{0mm}{4mm}
  \smtick{CRP} & 0.0730 & 0.0908 & 0.0773 & 0.0817 & 0.0862 & 0.0687 & 0.0774 \\
  \hline 
\end{array}
$
\end{table}\vspace*{-1ex}%
with the corresponding
\href{https://en.wikipedia.org/wiki/PGF/TikZ}{TikZ} image
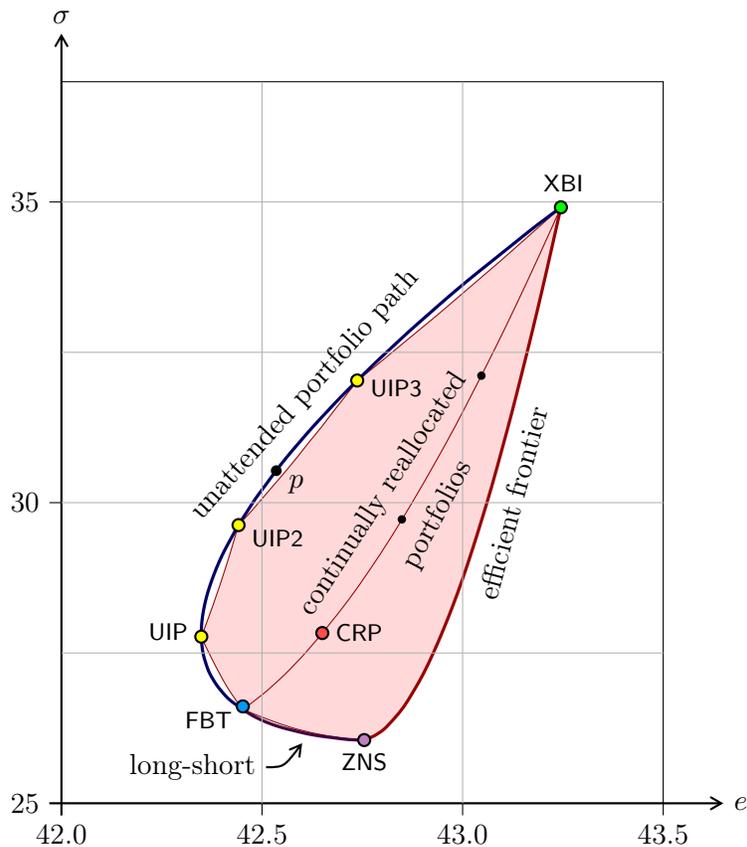
\begin{figure}[H]
  \centering
  \caption{\label{last-look}
  How not to do MV-analysis -- a last look}
  \vspace*{-1ex}
\begin{tikzpicture}[scale=0.8,>={angle 60}]
  \setlength{\pointradius}{0.10cm}
  \setlength{\smallradius}{0.7\pointradius}
  
  \colorlet{unattended}{blue!50!black}  
  \colorlet{reallocated}{red!50!black}  
  \colorlet{CRi}{red!15}  
  
  \newcommand{\xL}{0}  
  \newcommand{\xH}{10}  
  \newcommand{\dx}{3.33333}  
  \newcommand{\yL}{25}      
  \newcommand{\yH}{37}      
  \newcommand{\dy}{5}
  
  \coordinate (FBT)  at (3.014,26.61);
  \coordinate (XBI)  at (8.300,34.91);
  \coordinate (UIP)  at (2.324,27.77);
  \coordinate (UIP2) at (2.943,29.625);
  \coordinate (UIP3) at (4.915,32.03);
  \coordinate (ZNS)  at (5.031,26.05);
  \coordinate (CRP)  at (4.335,27.83);  
  \coordinate (CRP2) at (5.657,29.72);
  \coordinate (CRP3) at (6.978,32.11);
  \coordinate (P)    at (3.570,30.53);
  
  \draw (\xL,\yL) rectangle (\xH,\yH);
  \draw (\xL,\yL) -- (\xL,\yL-0.2) node[below]{42.0};
  \draw (\xL+\dx,\yL) -- (\xL+\dx,\yL-0.2) node[below]{42.5};
  \draw (\xL+2*\dx,\yL) -- (\xL+2*\dx,\yL-0.2) node[below]{43.0};
  \draw (\xH,\yL) -- (\xH,\yL-0.2) node[below]{43.5};
  \draw (\xL,\yL) -- (\xL-0.2,\yL) node[left]{25};
  \draw (\xL,\yL+5) -- (\xL-0.2,\yL+5) node[left]{30};
  \draw (\xL,\yL+10) -- (\xL-0.2,\yL+10) node[left]{35};
  
  \draw[thick,->] (\xL-0.2,\yL) -- (\xH+1,\yL) node[right]{$e$};
  \draw[thick,->] (\xL,\yL-0.2) -- (\xL,\yH+0.8) node[above]{$\sigma$};
  
  \filldraw[fill=CRi,draw=CR] plot[smooth] coordinates {
  (5.031,26.05) 
  (5.195,26.12) (5.358,26.23) (5.521,26.40) (5.685,26.60) (5.848,26.86)
  (6.012,27.15) (6.175,27.49) (6.339,27.87) (6.502,28.28) (6.666,28.73)
  (6.829,29.22) (6.992,29.74) (7.156,30.29) (7.319,30.88) (7.483,31.49)
  (7.646,32.13) (7.810,32.79) (7.973,33.47) (8.136,34.18) (8.300,34.91)
  } -- plot[smooth] coordinates {
  (8.300,34.91) 
  (7.961,34.61) (7.623,34.31) (7.284,34.01) (6.946,33.72) (6.608,33.43)
  (6.269,33.14) (5.931,32.86) (5.592,32.58) (5.254,32.30) (4.915,32.03)
  } -- plot[smooth] coordinates {
  (4.915,32.03) 
  (4.718,31.77) (4.521,31.52) (4.323,31.26) (4.126,31.02) (3.929,30.77)
  (3.732,30.53) (3.535,30.30) (3.337,30.07) (3.140,29.84) (2.943,29.63)
  } -- plot[smooth] coordinates {
  (2.943,29.63) 
  (2.881,29.42) (2.819,29.21) (2.757,29.01) (2.696,28.82) (2.634,28.63)
  (2.572,28.45) (2.510,28.27) (2.448,28.10) (2.386,27.93) (2.324,27.77)
  } -- plot[smooth] coordinates {
  (2.324,27.77) 
  (2.393,27.62) (2.462,27.48) (2.531,27.34) (2.600,27.21) (2.669,27.08)
  (2.738,26.96) (2.807,26.85) (2.876,26.75) (2.945,26.65) (3.014,26.56)
  } -- plot[smooth] coordinates {
  (3.014,26.56) 
  (3.417,26.40) (3.821,26.26) (4.224,26.16) (4.628,26.09) (5.031,26.05)
  };
  \draw[CR, very thick]  plot[smooth] coordinates {
  (5.031,26.05) 
  (5.195,26.12) (5.358,26.23) (5.521,26.40) (5.685,26.60) (5.848,26.86)
  (6.012,27.15) (6.175,27.49) (6.339,27.87) (6.502,28.28) (6.666,28.73)
  (6.829,29.22) (6.992,29.74) (7.156,30.29) (7.319,30.88) (7.483,31.49)
  (7.646,32.13) (7.810,32.79) (7.973,33.47) (8.136,34.18) (8.300,34.91)
  };
  
  \draw[CR] plot[smooth] coordinates {
  (3.014,26.56) 
  (3.366,26.83) (3.719,27.15) (4.071,27.52) (4.423,27.94) (4.776,28.39)
  (5.128,28.89) (5.481,29.43) (5.833,30.01) (6.185,30.62) (6.538,31.26)
  (6.890,31.94) (7.243,32.64) (7.595,33.37) (7.948,34.13) (8.300,34.91)
  };
  
  \draw[FZ,very thick] plot[smooth] coordinates {
  (5.031,26.05) 
  (4.770,26.06) (4.523,26.09) (4.288,26.12) (4.067,26.16) (3.858,26.21)
  (3.663,26.26) (3.481,26.32) (3.312,26.39) (3.156,26.47) (3.014,26.56)
  };
  \draw[XF,very thick] plot[smooth] coordinates {
  (3.014,26.56) 
  (2.887,26.65) (2.773,26.74) (2.672,26.85) (2.584,26.96) (2.508,27.08)
  (2.445,27.20) (2.396,27.34) (2.359,27.47) (2.335,27.62) (2.324,27.77)
  (2.327,27.93) (2.342,28.09) (2.371,28.26) (2.413,28.44) (2.468,28.62)
  (2.536,28.81) (2.618,29.01) (2.713,29.21) (2.821,29.41) (2.943,29.63)
  (3.078,29.84) (3.227,30.07) (3.390,30.29) (3.566,30.53) (3.756,30.77)
  (3.960,31.01) (4.178,31.26) (4.410,31.51) (4.655,31.77) (4.915,32.03)
  (5.189,32.30) (5.477,32.57) (5.779,32.85) (6.096,33.13) (6.427,33.42)
  (6.772,33.71) (7.132,34.00) (7.507,34.30) (7.896,34.61) (8.300,34.91)
  };

  \draw[grid] (\dx,\yL) -- (\dx,\yH);
  \draw[grid] (2*\dx,\yL) -- (2*\dx,\yH);
  \draw[grid] (\xL,27.5) -- (\xH,27.5);
  \draw[grid] (\xL,30) -- (\xH,30);
  \draw[grid] (\xL,32.5) -- (\xH,32.5);
  \draw[grid] (\xL,35) -- (\xH,35);

  \node[rotate=57] at (5.35,30.15) {continually reallocated};
  \node[rotate=62] at (6.25,29.8) {portfolios};
  \node[rotate=75] at (7.5,30) {efficient frontier};
  \node[rotate=48] at (4.05,31.8) {unattended portfolio path};
  \node[left] at (3.4, 25.6) {long-short};
  \draw[thick, ->] (3.4,25.6) .. 
    controls (3.7,25.6) and (3.7,25.6) .. (4.0,26.0);
  
  \filldraw[fill=FBTv,thick] (FBT) circle[radius=\pointradius]
    node[below=5,left]{\smtick{FBT}};
  \filldraw[fill=XBIv,thick] (XBI) circle[radius=\pointradius]
    node[right=1,above=2]{\smtick{XBI}};
  \filldraw[fill=UIPv,thick] (UIP) circle[radius=\pointradius]
    node[above=2,left=1]{\smtick{UIP}};
  \filldraw[fill=UIPv,thick] (UIP2) circle[radius=\pointradius]
    node[below=5,right=1]{\smtick{UIP2}};
  \filldraw[fill=UIPv,thick] (UIP3) circle[radius=\pointradius]
    node[below=3,right=1]{\smtick{UIP3}};
  \filldraw[fill=ZNSv,thick] (ZNS) circle[radius=\pointradius]
    node[below=1]{\smtick{ZNS}};
  \filldraw[fill=CRPv,thick] (CRP) circle[radius=\pointradius]
    node[above=0,right=1]{\smtick{CRP}};
  \fill[black] (CRP2) circle[radius=\smallradius];
  \fill[black] (CRP3) circle[radius=\smallradius];
  \fill[black] (P) circle[radius=0.9\pointradius] node[below=5,right=1]{$p$};
\end{tikzpicture}
\end{figure}
\vspace*{-1ex}%
Here the continually-reallocated black-dotted portfolio points are
exactly 1/2 and 3/4 of the $e$-way from \ticker{FBT} to \ticker{XBI} on
the red, continually-reallocated, \ticker{FBT}-to-\ticker{XBI} path.

\newpage
\section{Conclusion}\label{conclusion}

The growth in value of an unattended investment portfolio \ticker{P}
 over a given
interval of time can be completely described by a normalized adjusted
closing price equation
\begin{equation}\label{adj_equation}
  \mathbf{a}_{\sstick{P}} = \sum_j p_j \mathbf{a}_j,\\[-1ex]
\end{equation}
where the $p_j$ are the proportions of the securities in the portfolio
 \ticker{P} at the close of the day of normalization.
The corresponding mean periodic return equation,
\begin{equation}\label{mean_equation}
  e_{\sstick{P}} = \sum_j p_j e_j,\\[-1ex]
\end{equation}
\emph{does not} follow when $e = \verb!mean!(\mathbf{r})$ and periodic
return vectors $\mathbf{r}$ are defined by
\begin{equation}\label{periodic_returns}
  r_i = a_i / a_{i-1} - 1.
\end{equation}

The mean periodic return equation \eqref{mean_equation} does hold with
\eqref{periodic_returns}  when one restricts his attention to
continually reallocated portfolios. Unfortunately continually
reallocated investment portfolios are more numerical artifact than
financial reality.

\appendix
\newpage
\section{An adjusted closing price primer}
\label{adjclose_primer}

The adjusted closing prices of a security are artificial ``closing
prices'' that are adjusted to incorporate all dividends and splits. The
day-to-day growth of a security or an unattended investment portfolio of
securities is completely described by its adjusted closing prices. If
the adjusted closing price of the security/portfolio is $a_0$ on market
day 0 and $a_1$ on a later market day 1, then its total return from day
0 to day 1 is ~$r = \Delta{a}/a_0 ~(\Delta{a} = a_1 - a_0)$. Two
adjusted closing price vectors for a given security that cover the same
time interval must be positive scalar multiples of each other. Thus the
returns, $r$, of the security from one market day to another do not
depend on any particular adjusted closing price representation.

Table \ref{generate_XBI_adjclose} shows how one can compute adjusted
closing prices for the exchange traded fund \ticker{XBI} over the period
2013-12-31 through 2014-12-31. The required input data are all closing
prices for the fund over this period as well as the dividends it made
during the period with their ex-dividend dates. On each line the adjusted closing
price is computed by\\\hspace*{2cm}%
$ \text{adjusted closing price}~=~\text{closing price}~\times
  ~\text{adjusted closing shares}\,.$\\[-1ex]

The adjusted closing shares in the table increase on each
ex-dividend day and are constant in between. If the closing price on the
market day prior to an ex-dividend day is $c_0$ and the dividend on the
ex-dividend day is $d_1$, then the adjusted closing shares on the
ex-dividend day must be increased by a factor of $c_0 / (c_0 - d_1)$ in
order that an investor who has his dividends reinvested maintains the
value of his investment.\\[-1ex]

The adjusted closing prices in Table \ref{generate_XBI_adjclose} are
``normalized'' at 100.000 on 2013-12-31. To compute the adjusted closing
prices for the 243 missing days just fill in the missing closing prices
and multiply them by the corresponding adjusted closing shares. Also
note that these closing prices and distributions have \emph{not} been
adjusted for the 3:1 split in 2015.

\begin{table}[H]
  \centering
  \captionsetup{width=9.5cm}
  \caption{\label{generate_XBI_adjclose}%
    How to generate normalized adjusted closing prices
    for \ticker{XBI} -- SPDR S\&P Biotech ETF
  }
\begin{tabular}{|c|rrrr|}\hline\rule{0mm}{4mm}
  & & & adjusted & adjusted \\
  market & distri- & closing & closing & closing \\
  day & bution & price & shares & price\rule[-2mm]{0mm}{2mm} \\\hline
  2013-12-31 & & 130.20 & 0.768049 & 100.000\rule{0mm}{3.8mm} \\
  $\cdots$ & & $\cdots$~~ & 0.768049 & $\cdots$~~ \\
  2014-03-20 & & 160.17 & 0.768049 & 123.018 \\
  2014-03-21 & 0.333023 & 153.15 & 0.769649 & 117.872 \\
  $\cdots$ & & $\cdots$~~ & 0.769649 & $\cdots$~~ \\
  2014-06-19 & & 153.32 & 0.769649 & 118.003 \\
  2014-06-20 & 0.616142 & 153.42 & 0.772754 & 118.556 \\
  $\cdots$ & & $\cdots$~~ & 0.772754 & $\cdots$~~ \\
  2014-09-18 & & 159.94 & 0.772754 & 123.594 \\
  2014-09-19 & 0.562774 & 158.27 & 0.775483 & 122.736 \\
  $\cdots$ & & $\cdots$~~ & 0.775483 & $\cdots$~~ \\
  2014-12-18 & & 189.08 & 0.775483 & 146.628 \\
  2014-12-19 & 0.490997 & 190.34 & 0.777502 & 147.990 \\
  $\cdots$ & & $\cdots$~~ & 0.777502 & $\cdots$~~ \\
  2014-12-31 & & 186.46 & 0.777502 & 144.973 \\\hline
\end{tabular}
\end{table}

\subsection{Yahoo!Finance}\label{yahoo}

\href{https://finance.yahoo.com}{Yahoo!Finance} is a good source for
adjusted closing prices of an individual security, like our
\href{https://finance.yahoo.com/quote/XBI/history?p=XBI}{\ticker{XBI}}.
Simply download
the daily, historical prices over the time interval desired as a
CSV (comma-separated-value) file and open the file in a spreadsheet
program (e.g., Excel).

This spreadsheet will have seven labeled columns:\\
\hspace*{5ex}Date,Open,High,Low,Close,Adj Close,Volume.\\
Delete all but the ``Date'' and the ``Adj Close'' columns. We will
assume these are now columns A and B, respectively, as in the
spreadsheet image (Figure
\ref{yahoo_XBI}) below.

The Yahoo adjusted closing prices are not normalized at any particular
date. Yahoo simply sets the adjusted closing price of a security at the
close of the latest market day equal to its closing price on that day,
Then previous adjusted closing prices must be rescaled if the day
is an ex-distribution or an ex-split day.

To normalize your Yahoo adjusted prices at say 100 on a particular date
(i.e., 2013-12-31 in Figure \ref{yahoo_XBI}) start a new normalized
adjusted closing price column on your spreadsheet, say column C, by
putting
\begin{equation}\label{B_date_row}
  \text{\fcolorbox{green!60!black}{green!10}{= B[date row]*100/B\$[date row]}
\hspace*{6ex}(with [date row] = the row number)}
\end{equation}
in the cell of that date. The number 100 should appear in this cell
(i.e., in cell C2). Now you need only fill up and/or down from this cell
to get all normalized adjusted closing prices in column C. (We only filled
down in in Figure \ref{yahoo_XBI}.)

\begin{figure}[H]
  \centering
  \caption{\label{yahoo_XBI}%
    \ticker{XBI} data from Yahoo on 2018-09-19
  }
  \includegraphics[scale=0.85]{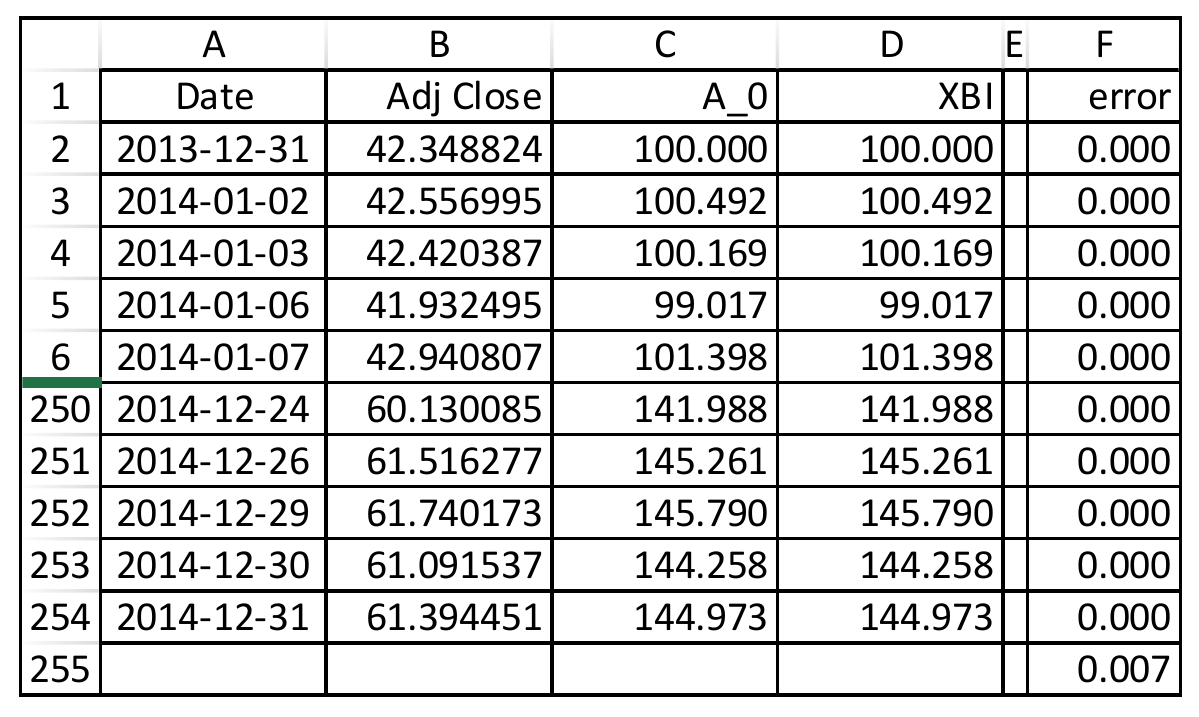}
\end{figure}

The normalized adjusted closing prices in column C of Figure
\ref{yahoo_XBI} were generated from the prices in column B as described
above. These prices were then rounded to 3 decimal places. The numbers
in the D column come from our \verb!anc/FXUZ7.cvs! file. They were
generated by the process used to generate Table
\ref{generate_XBI_adjclose} from the closing prices and distributions of
\ticker{XBI}. Out of the 253 market days considered, the C
(\ticker{A\_0}) price was 0.001 greater than the D (\ticker{XBI}) price
on 7 days. Otherwise the two columns of adjusted closing prices were
exactly the same. (These 7 ``errors'' occurred in the 243 rows that have
been collapsed in Figure \ref{yahoo_XBI}.)

\newpage
\section{Life on the Mississippi}
\label{twain}

In the space of one hundred and seventy-six years the Lower Mississippi
has shortened itself two hundred and forty-two miles. That is an average
of a trifle over one mile and a third per year. Therefore, any calm
person, who is not blind or idiotic, can see that in the Old Oolitic
Silurian Period, just a million years ago next November, the Lower
Mississippi River was upwards of one million three hundred thousand
miles long, and stuck out over the Gulf of Mexico like a fishing-rod.
And by the same token any person can see that seven hundred and
forty-two years from now the Lower Mississippi will be only a mile and
three-quarters long, and Cairo and New Orleans will have joined their
streets together, and be plodding comfortably along under a single mayor
and a mutual board of aldermen. There is something fascinating about
science. One gets such wholesale returns of conjecture out of such a
trifling investment of fact.

\hspace*{10cm}-- ~Mark Twain

\clearpage
\printbibliography  

\end{document}